\documentclass[fleqn,usenatbib]{mnras}

\usepackage[T1]{fontenc}
\usepackage{ae,aecompl}

\usepackage{graphicx}	
\usepackage{amsmath}	
\usepackage{amssymb}	
\usepackage{pdflscape}
\usepackage{times,txfonts}
\usepackage{hyperref}	

\newcommand{\feh}{\ensuremath{\mathrm{[Fe/H]}}}
\newcommand{\afe}{\ensuremath{\mathrm{[\alpha/Fe]}}}
\newcommand{\mgfe}{\ensuremath{\mathrm{[Mg/Fe]}}}
\def\red{}

\title[Weighing the stellar halo]{Weighing the stellar constituents of the Galactic halo with APOGEE red giant stars}

\author[J. T. Mackereth et al.]{
J. Ted Mackereth$^{1}$\thanks{E-mail: j.e.mackereth@bham.ac.uk (UoB)}
and Jo Bovy$^{2}$
\\
$^{1}$School of Astronomy and Astrophysics, University of Birmingham, Edgbaston, Birmimgham, B15 2TT, UK\\
$^{2}$Department of Astronomy and Astrophysics, University of Toronto, 50 St. George Street, Toronto, ON M5S 3H4, Canada\\
}

\date{}

\pubyear{2019}

\begin{document}
\label{firstpage}
\pagerange{\pageref{firstpage}--\pageref{lastpage}}
\maketitle

\begin{abstract}
The stellar mass in the halo of the Milky Way is notoriously difficult to determine, owing to the paucity of its stars in the solar neighbourhood. With tentative evidence from \emph{Gaia} that the nearby stellar halo is dominated by a massive accretion event -- referred to as \emph{Gaia-Enceladus} or Sausage -- these constraints are now increasingly urgent. We measure the mass in kinematically selected mono-abundance populations (MAPs) of the stellar halo between $-3 <$ [Fe/H] $< -1$ and $0.0 <$ [Mg/Fe] $< 0.4$ using red giant star counts from APOGEE DR14. We find that MAPs are well fit by single power laws on triaxial ellipsoidal surfaces, and we show that that the power law slope $\alpha$ changes such that high [Mg/Fe] populations have $\alpha \sim 4$, whereas low [Mg/Fe] MAPs are more extended with shallow slopes, $\alpha \sim 2$. We estimate the total stellar mass to be $M_{*,\mathrm{tot}} = 1.3^{+0.3}_{-0.2}\times10^{9}\ \mathrm{M_{\odot}}$, of which we estimate $\sim 0.9^{+0.2}_{-0.1} \times 10^{9} \mathrm{M_{\odot}}$ to be accreted. We estimate that the mass of accreted stars with $e > 0.7$ is $M_{*,\mathrm{accreted},e>0.7} = 3 \pm 1\ \mathrm{(stat.)}\pm 1\ \mathrm{(syst.)}\times10^{8}\ \mathrm{M_\odot}$, or $\sim 30-50\%$ of the accreted halo mass. If the majority of these stars \emph{are} the progeny of a massive accreted dwarf, this places an upper limit on its stellar mass, and implies a halo mass for the progenitor of $\sim 10^{10.2\pm0.2}\  \mathrm{M_\odot}$. This constraint not only shows that the \emph{Gaia-Enceladus}/Sausage progenitor may not be as massive as originally suggested, but that the majority of the Milky Way stellar halo was accreted. These measurements are an important step towards fully reconstructing the assembly history of the Milky Way. 
\end{abstract}




\section{Introduction}

The stellar haloes of galaxies host graveyards of debris left behind by accreted satellites which have contributed to the assembly of their mass, containing important information about their early formation and evolution. Stellar haloes are widely found to contain clear substructure in the form of streams and shells \citep[e.g.][]{2008ApJ...689..184M,2015AJ....150..116M}, which provide a means for understanding galaxy formation \citep[e.g.][]{2005ApJ...635..931B}, and testing models for dark matter \citep[e.g.][]{2016ApJ...833...31B,2017MNRAS.466..628B}. However, it is the mass contained in the smoother components of stellar haloes, and its distribution in terms of stellar element abundances and kinematics, which can provide insight into their early assembly, and the merging events which have occurred, whose debris is by now smoothly mixed in the haloes.

The Milky Way stellar halo shape and mass profile is well studied using large samples of stars, generally concentrating on bright tracers of its entire stellar population which have easily attainable distance measures such as RR Lyrae \citep[e.g.][]{2019MNRAS.482.3868I,2013AJ....146...21S}, Main Sequence Turn-Off (MSTO) stars \citep[e.g][]{2011ApJ...731....4S}, and Blue Horizontal Branch (BHB) stars \citep[e.g.][]{2011MNRAS.416.2903D}. Such studies generally find that the MW halo is best described as having a broken density profile \citep[interestingly, M31 appears to have the opposite, e.g.][]{2013ApJ...763..113D} which has a power law index between $\sim 2$ and $\sim 4$ depending on the radius studied. The nature of these tracers means that they are subject to certain biases in age or metallicity, but a few studies have looked at Red Giant Branch (RGB) stars \citep{2015ApJ...809..144X} and used full CMD fitting \citep{2010ApJ...714..663D}, avoiding much of the biases, while finding similar results. 

Inner and outer halo components are also reflected in the metallicities of halo stars \citep{2007Natur.450.1020C}, which may or may not reflect the presence of the \emph{in} and \emph{ex situ} components of the stellar halo \citep[see evidence for the opposite in, e.g.][]{2011MNRAS.415.3807S,2014ApJ...786....7S}. The scarcity of extensive spectroscopic data has thus far meant that it has not been possible to easily divide and study the halo in terms of element abundances and kinematics, which may shed light on this apparent `dual' nature of the halo. Cosmological numerical simulations seem to suggest that such duality arising from \emph{in} and \emph{ex situ} halo components is plausible at least \citep[e.g.][]{2011MNRAS.416.2802F,2012MNRAS.420.2245M}. Another paper based on simulated haloes by \citet{2013ApJ...763..113D} suggested that the broken density profile may be attributed to a single massive accretion event, whose stars were piled up at the apocenter of its orbit. Combining spectroscopic data with positions and kinematics on larger scales will elucidate these connections.

Aside from the density slope and structure of the halo, the question of its stellar mass content is also an important one, and similarly is affected by the difficulty in correcting the counts of tracer populations to the entire underlying halo. Estimates of the mass are immutably connected to the derived mass normalisation per tracer star, which generally relies on stellar models and intricate calibrations. A recent pre-print of \citet{2019arXiv190802763D} compiled estimates from the literature, ranging between $3$ and $15\times10^{-5}\ \mathrm{M_{\odot}\ pc^{-3}}$ \citep[e.g.][]{1993AJ....106..578M,1998A&A...329...81F,1998ApJ...503..798G,2003MNRAS.344..583D,2008ApJ...673..864J,2010ApJ...714..663D}, with that work finding $8.1$ or $6.9\times10^{-5}\ \mathrm{M_{\odot}\ pc^{-3}}$, depending on whether or not the Sagittarius dwarf is included, respectively. The total stellar mass has been found to be somewhere in the region of a few $10^{8} \mathrm{M_{\odot}}$ \citep[e.g.][and references therein]{2018arXiv180902658B},  but the recent \citet{2019arXiv190802763D} pre-print suggests a much higher value of $1.4\pm0.4\times10^{9}\ \mathrm{M_{\odot}}$, based on red giant number counts and extrapolation of the \citet{1965TrAlm...5...87E} profile fit previously by \citet{2011MNRAS.416.2903D}. 

Related to the possibility of a high stellar mass of the halo, the recent advent of the \emph{Gaia} DR2 data has revealed a number of likely remnants of dwarf galaxies which merged with the Milky Way early in its history. A number of groups have suggested the existence of at least one \citep[e.g.][]{2018MNRAS.478..611B,2018arXiv180606038H}, if not two \citep[e.g.][]{2019ApJ...870L..24B,2019MNRAS.488.1235M} possible progenitor dwarfs, which can be disentangled from the rest of the halo on the basis of their kinematics. \citet{2018arXiv180800968M} showed that the high eccentricity halo component \citep[related to the larger of the discovered remnants, \emph{Gaia-Enceladus} or, the Sausage, e.g.][]{2018arXiv180606038H,2018MNRAS.478..611B} is distinct in element abundances \citep[see also][]{2019arXiv190309320D}, and studied the occurence of such merger events in the EAGLE simulations \citep[][]{2015MNRAS.446..521S,2015MNRAS.450.1937C}, showing that high $e$ debris is rare and arises from later merging events.  \citet{2019MNRAS.484.4471F} also study such merger events in cosmological zoom-in simulations, finding that they invoke rapid increases in the galaxy halo mass. Similar connections between the MW disc abundances and rapid growth of its halo have also been made on the basis of cosmological simulations \citep{2018MNRAS.477.5072M}. Finally, a third, significantly massive (and hitherto unaccounted for) dwarf, dubbed \emph{Kraken}, is also indicated by the age-metallicity distribution of the Galactic globular clusters \citep[e.g.][]{2018MNRAS.tmp.1537K}. 

Regardless of the exact division between these large dwarf remnants, which is still to be exactly determined, it seems at least that the large majority of high eccentricity ($e > 0.7$) halo stars are likely to be accreted. Mass estimates for this high $e$ accreted component have ranged in recent literature from a few $10^8\ \mathrm{M_{\odot}}$ to as high as a few $10^{9}\ \mathrm{M_{\odot}}$ \citep[e.g.][]{2018MNRAS.478..611B,2019arXiv190309320D,2018arXiv180606038H,2018arXiv180800968M,2019MNRAS.487L..47V}, with little consensus on how this can be reconciled with the literature stellar halo total mass, other than to say that it is likely this component makes up a large fraction of it ($> 50\%$). The \cite{2019arXiv190802763D} work concludes that their high halo mass is indicative that such a large merger did occur, when compared with similar galaxies in the AURIGA simulations \citep{2017MNRAS.467..179G}. A direct estimate of the mass contained in the high eccentricity halo would thus be a useful point on the trace of the mass assembly history of the Galaxy. 

In this paper, we present a novel mass estimate for the stellar halo of the Milky Way, obtrained via density modelling of APOGEE red giant stars, selected on the basis of their element abundances and kinematics. Our methodology allows for the assessment of the stellar mass contained in the high and low eccentricity stellar halo, as well as studying the mass distribution as a function of \feh{} and \mgfe{}. We demonstrate that reliable models for the stellar density can be fit to the data with remarkably low numbers of stars. The APOGEE red giant star counts are relatively simply corrected for the halo normalisation, through reconstruction of the APOGEE selection function and using stellar evolution models, allowing us to gauge the stellar halo mass to good precision and accuracy. 

The paper is laid out as follows: Section \ref{sec:data} presents our selection of APOGEE red giant stars, discussing the derivation of their orbital parameters using \emph{Gaia} DR2 data. In Section \ref{sec:method} we reformulate the density modelling procedure of \citet{2016ApJ...823...30B} and \citet{2017arXiv170600018M}, including the allowances for the APOGEE-2 selection function and for modelling of the full 6D phase space density. Section \ref{sec:results} presents the main results of the paper, which are discussed in the context of the previous work on the halo in Section \ref{sec:discussion}. Finally, in Section \ref{sec:conclusions}, we summarise our results and provide concluding statements.

\begin{figure}
\centering
\includegraphics[width=0.9\columnwidth]{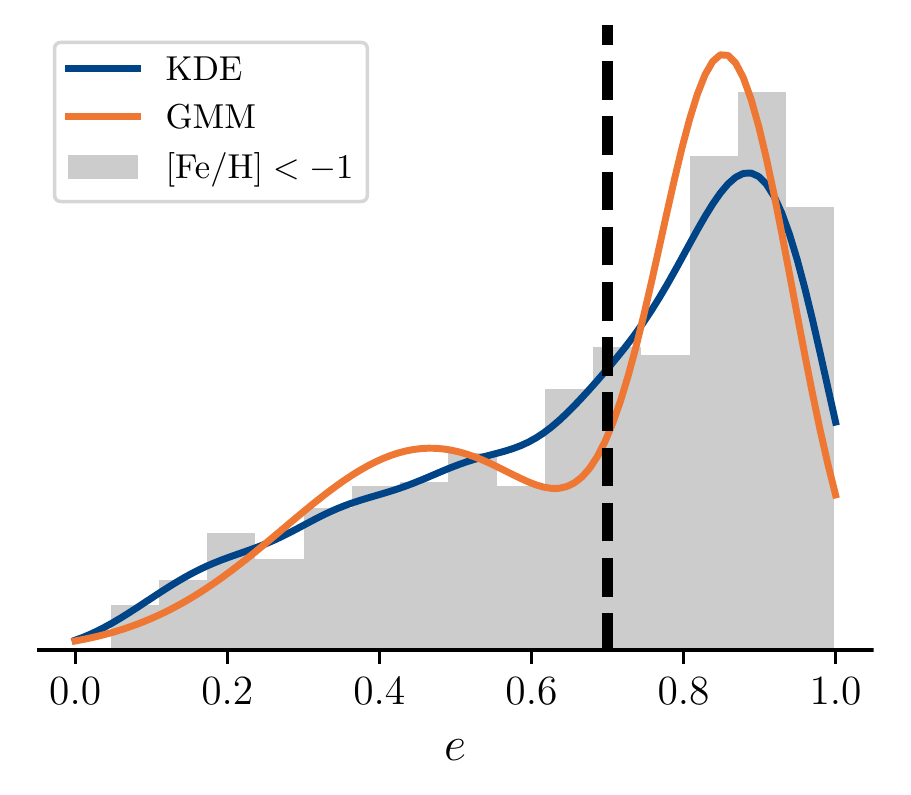}
\caption{\label{fig:eccdist} The eccentricity distribution of the \feh{} $<-1.$ sample. The gray histogram shows the raw star counts, and the overplotted lines show different models of the PDF of the observed distribution. The blue line shows a simple Gaussian kernel density estimate, with a kernel width determined via the Silverman method. The orange curve demonstrates the best fit 1D two component Gaussian mixture model. A GMM with two components describes the data well, and provides a well motivated division in eccentricity at $e = 0.7$, which is the eccentricity at which the probability of a star being in either component is equal.}
\end{figure}

\begin{figure}
\includegraphics[width=\columnwidth]{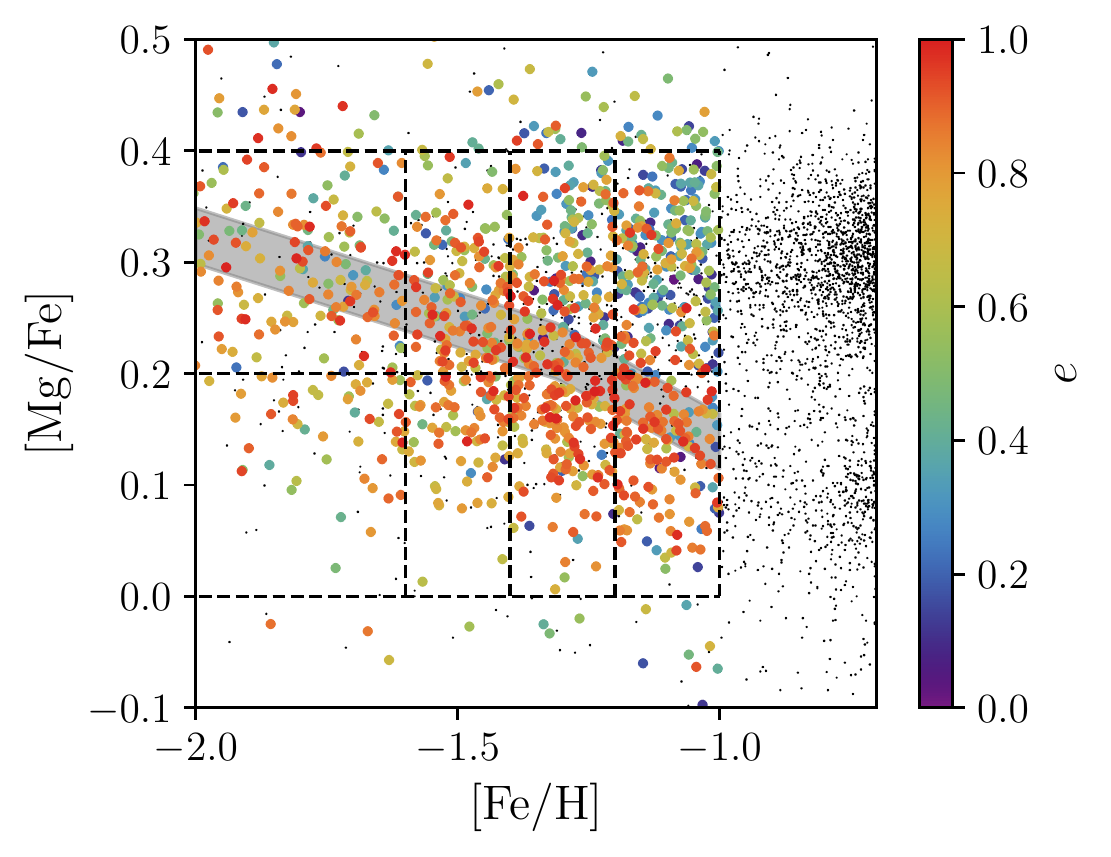}
\caption{\label{fig:afe_feh_sample} The \afe{}-\feh{} distribution of APOGEE DR14 red giants at low \feh{}. Stars with \feh{} $< -1$ which are in our sample of giants are coloured according to their orbital eccentricity $e$, and are plotted in order of increasing $e$. The overlaid dashed grid shows the adopted MAPs used in our analysis, having $\Delta\mathrm{[Fe/H]} = 0.2$ dex and $\Delta\mathrm{[Mg/Fe]} = 0.2$ dex. The two MAPs (shown by dotted lines) at the lowest \feh{} extend down to $\feh{} = -3$ in order to capture the star counts at very low \feh{}, where the abundance uncertainty becomes high. A population of mid-to-high eccentricity debris identified previously corresponding to the \emph{Gaia}-Sausage and Sequoia \citep[e.g.][]{2019MNRAS.488.1235M} or \emph{Gaia-Enceladus} \citep[e.g.][]{2018arXiv180606038H} is clearly visible running through the lower \mgfe{} bins, and is indicated here by a gray band.}
\end{figure}

\section{Data}
\label{sec:data}
We use data from the fourteenth data release \citep[DR14][]{2018ApJS..235...42A} of the SDSS-IV Apache Point Observatory Galactic Evolution Experiment \citep[APOGEE,][]{2015arXiv150905420M}. APOGEE is a spectroscopic survey of stars in the Milky Way, and refers collectively to the APOGEE-1 and 2 surveys, conducted under SDSS-III \citep{2011AJ....142...72E} and SDSS-IV \citep{2017AJ....154...28B}, respectively. Conducted in the near infra-red (NIR) H-Band, it has observed high signal-to-noise ratio (SNR $> 100\ \mathrm{pixel^{-1}}$), high resolution ($R \sim 20,000$) spectra for well over $2\times10^{5}$ stars in the Milky Way, across a large range of Galactocentric radii, and heights above the Galactic mid-plane. Spectra are observed using the APOGEE spectrograph \citep{2019arXiv190200928W}, fed from the SDSS 2.5m telescope at APO \citep{2006AJ....131.2332G}. Data are reduced using the APOGEE data reduction pipeline \citep{2015AJ....150..173N}. Throughout the paper, we use the element abundances which are generated by the APOGEE Stellar Parameters and Chemical Abundances Pipeline \citep[ASPCAP,][]{2016AJ....151..144G}. Abundances are measured using a specially computed linelist \citep{2015ApJS..221...24S}, a pre-computed library of synthetic stellar spectra \citep{2015AJ....149..181Z}. In DR14, stellar parameters, 19 element abundances and heliocentric radial velocities are provided \citep{holtzdr14}, and are tested rigourously against literature standards \citep{jonssondr14}. Stars for APOGEE-1 and 2 were selected from the 2MASS catalogue \citep{2006AJ....131.1163S}, using a target selection scheme outlined in \citep{2013AJ....146...81Z,2017AJ....154..198Z}. The reproduction of the survey selection function is necessary to properly model APOGEE star counts and has been described in \citet{2014ApJ...790..127B}. Here, we extend that work to account for the additional selection criteria on APOGEE-2 targets, and fully describe the procedure in Appendix \ref{sec:appA}. 

We adopt the artificial neural network (ANN) based distance estimates from \citet{2019arXiv190208634L}, which were generated using the \texttt{astroNN} python package \citep[first described in][]{2018arXiv180804428L}. The method and the specific ANN architecture is described in detail by \citet{2019arXiv190208634L}. In brief, the ANN used is trained to predict stellar luminosity from infra-red spectra using a training set comprising of stars with APOGEE spectra and \emph{Gaia} DR2 parallax measurements \citep{2018arXiv180409365G}. The model is able to simultaneously predict distances and account for the parallax offset present in \emph{Gaia}-DR2, producing high precision, accurate distance estimates for APOGEE stars, which match well with external catalogues and standard candles. Our methodology for estimating the mass of the stellar halo relies only on the 3D stellar positions, and we require data from \emph{Gaia} DR2 \citep{2018arXiv180409365G} purely for the stellar kinematics, which we use to define our sub-samples in eccentricity, described below.

We select a sample of red giant stars with well measured distances from APOGEE, accepting only stars with $1 < \log(g) < 3$, and whose distance uncertainty is less than 20\%. We consider all stars with $-3 < \feh{} < -1$ and $0.0 < \mgfe{} < 0.4$, and note that we use the catalogued ASPCAP abundances, rather than those from \texttt{astroNN} \citep{2018arXiv180804428L}. We also remove stars which are in fields towards the Galactic center between $20^{\circ} < l < 340^{\circ}$ and $|b| < 20^{\circ}$, and those which contained globular clusters used for calibration in APOGEE (in this case, only one field had stars in the statistical sample). Our final APOGEE sample is made up of 835 stars.

We perform a cross-match between our APOGEE sample and \emph{Gaia} DR2 \citep{2018arXiv180409365G} using the CDS X-match service\footnote{\url{http://cdsxmatch.u-strasbg.fr}}. Combining its proper motion measurements with APOGEE radial velocities and the \texttt{astroNN} distances, we determine full 6D phase-space coordinates for the stars in our sample. To convert between astrometric parameters and Galactocentric cylindrical coordinates, we assume a solar velocity computed by combining the proper motion of Sgr A* \citep{2018A&A...615L..15G} with the determination of the local standard of rest of \citet{2010MNRAS.403.1829S}, such that $\vec{v}_{\odot} = [U_{\odot}, V_{\odot}, W_{\odot}] = [-11.1,245.6,7.25]\ \mathrm{km\ s^{-1}}$. We assume the distance between the sun and the Galactic center to be $R_0=8.125\mathrm{kpc}$ \citep{2018A&A...615L..15G}, and the vertical height of the sun above the midplane $z_0 = 0.02\mathrm{kpc}$ of \citet{2018arXiv180903507B}. We then perform an estimation of the orbital eccentricities for 100 samples of the 6D ellipse defined by the uncertainty and covariance matrix of the observations of each star using the St\"ackel fudge method presented in \citet{2018arXiv180202592M} and implemented in \texttt{galpy} \citep{2015ApJS..216...29B}, assuming the simple \texttt{MWPotential2014} potential. We take the median eccentricity of the 100 samples for each star, and adopt this as their eccentricity. 

The resulting eccentricity distribution is shown in Figure \ref{fig:eccdist}. It is strongly peaked toward high $e$, with a median $e \sim 0.7$. The median uncertainty in our $e$ estimates is 0.1 across the full range of eccentricity (i.e. stars with higher eccentricity do not necessarily have larger uncertainties due to, for example, larger uncertainties in their proper motions). In order to assess the best point at which to divide our sample into two eccentricity bins, we make both a Gaussian kernel density estimate (KDE) of and fit a one dimensional, two component Gaussian mixture model (GMM) to the eccentricity distribution, shown in Figure \ref{fig:eccdist} as the blue and orange curves, respectively. We determine that the probability that a star belongs to either of the GMM components is equal at $e = 0.7$. This eccentricity also appears to correspond closely to a change in slope in the KDE curve. We therefore adopt this value as the division between high and low eccentricity sub-samples. It has already been shown in \citet{2018arXiv180800968M} that stars which are members of the proposed \emph{Gaia}-Sausage population are neatly divided from the rest of the halo by their orbit eccentricity (at a value in the region of $e\sim0.7$). 

We assess the level to which stars at large distances (and therefore large proper motion uncertainties) are deflected from the highest eccentricities into the $e < 0.7$ bin by checking the trend of eccentricity with distance for stars with abundances consistent with being members of \emph{Gaia-Enceladus}/Sausage (discussed further below, as shown in Figure \ref{fig:afe_feh_sample}). As we expect that these stars should all be nearly radial, the number of stars in this population at great distances with $e < 0.7$ gives an estimate of this fraction. We find that this number is low, and of the order $\sim 10\%$. The majority of stars which are affected by this effect are accounted for when we make an estimate of the systematic uncertainty on our mass estimate, in the discussion of the $e$ boundary, in Section \ref{sec:lowmass}.  

We select `mono-abundance' populations (MAPs) of halo stars in the low and high $e$ populations by making cuts in the \mgfe{}-\feh{} abundance space with $\Delta\mathrm{[Fe/H]} = 0.2$ dex and $\Delta\mathrm{[Mg/Fe]} = 0.2$ dex, as shown in Figure \ref{fig:afe_feh_sample}. {\red We select stars based on \mgfe{}, rather than a combination or a different selection of $\alpha$-elements because $\mathrm{Mg}$ provides the cleanest division between the \emph{Gaia-Enceladus}/Sausage population (and therefore likely other accreted populations) at these metallacities. \citet{jonssondr14} also demonstrate that $\mathrm{Mg}$ is the most accurately measured $\alpha$-element in APOGEE, relative to external measurements.} The sample used for this work is shown in that figure by the points coloured by their orbital eccentricity $e$ (gray points in the abundance range we consider are those which lie in bulge fields which are removed from our final sample, as discussed). It is clear from recent studies in the stellar halo that likely accreted stellar populations seem to dominate strongly among APOGEE stars at low \mgfe{} ($\lesssim 0.2$ dex) and at relatively low \feh{} \citep[$\lesssim -0.8$ dex, e.g.][]{2018arXiv180606038H,2018arXiv180800968M}, and we indicate these in Figure \ref{fig:afe_feh_sample} with a gray band. We group together all stars with $-3.0 < \mathrm{[Fe/H]} < -1.6$, dividing these only in \mgfe{}, as the number of stars in the sample becomes very low in this range of \feh{} and it is still pertinent to understand the contribution to the stellar mass of the halo of stars at these metallicities.

\begin{figure*}
\includegraphics[width=0.9\textwidth]{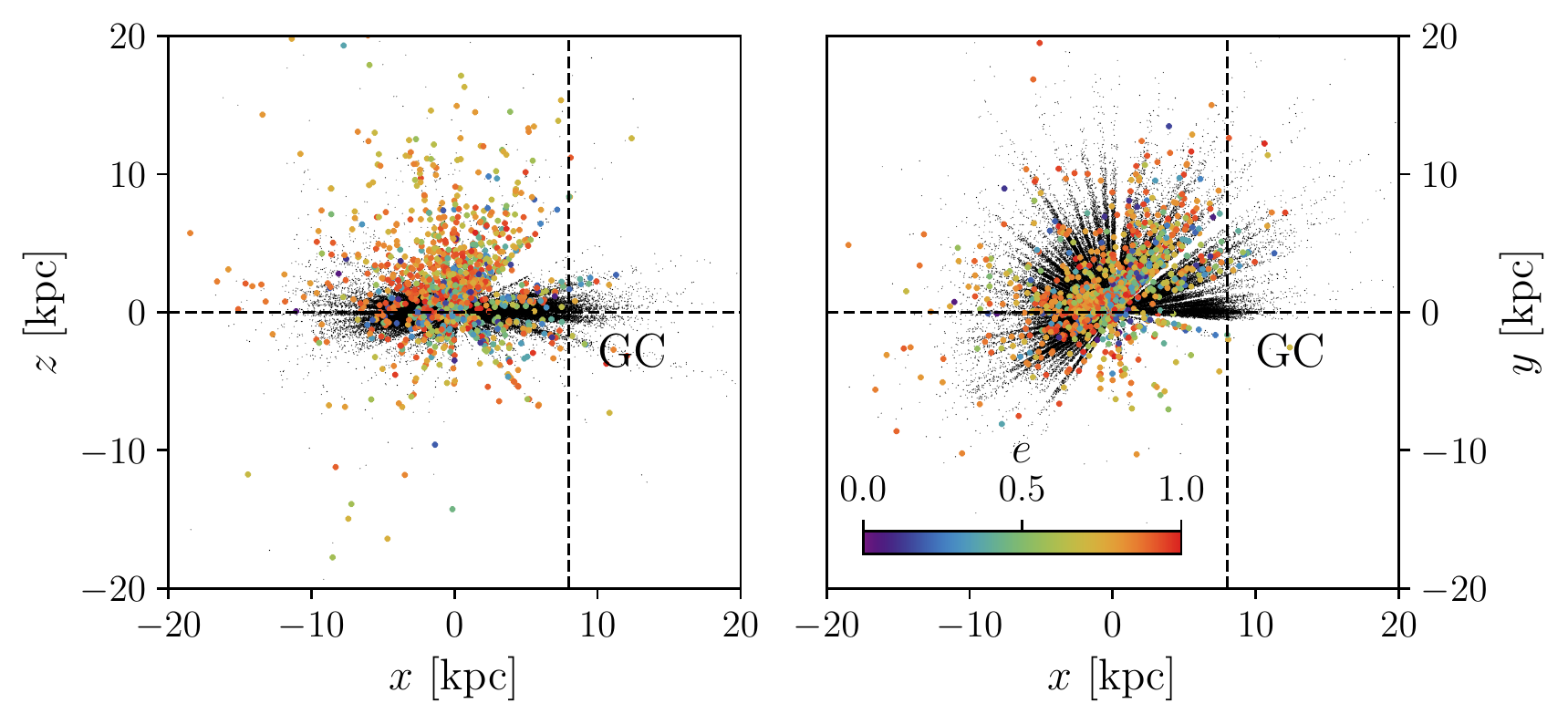}
\caption{\label{fig:spatial_sample} The spatial distribution in heliocentric $x$-$z$ and $x$-$y$ of our APOGEE DR14 sample, which is selected to prefer stars in the halo: all stars with $-3 < \mathrm{[Fe/H]} < -1$, and $0.0 < \mathrm{[Mg/Fe]} < 0.4$. The large points depict the sample, with colour describing the orbital eccentricity of the stars. The small black points demonstrate the spatial distribution of the entire APOGEE DR14 sample, which is mainly made up of disk stars. The crosshair marks the position of the Galactic centre. The coloured points are visibly more vertically and radially extended, even in the absence of any correction for the APOGEE selection function. APOGEE samples the halo well between $3< r_g < 13$ kpc, but the sample extends as far as $r_g \sim 50$ kpc for the most distant stars.}
\end{figure*}

Finally, we show the spatial distribution of the stars in each of the samples selected in \mgfe{}-\feh{} space (points, coloured by orbital eccentricity), compared to that of the whole APOGEE DR14 sample (black points), in Figure \ref{fig:spatial_sample}. Before any correction for the APOGEE selection function, or modelling of the stellar density, it is immediately obvious that the low \feh{} sample is less confined to the midplane of the Galaxy than the rest of APOGEE DR14. The stars in our selection sample well the region between $3 < r_g < 13$ kpc, but trace the halo as far as $r_g \sim 50$ kpc in the most extreme case. Since many studies suggest the halo density rapidly drops after $r_g \sim 20$ kpc \citep[e.g.][]{2011MNRAS.416.2903D}, and we fit a simple single power law, albeit with a cut-off, in the sampled region, it is possible that our final mass estimate may be a lower limit. However, we demonstrate later that our sample likely captures stellar populations which are part of both the shallow inner halo and the steeper outer halo.

\section{Method}
\label{sec:method}
\subsection{Modelling the density of kinematically selected MAPs}

We fit the density of giants in mono-abundance populations, accounting for the target selection of APOGEE and filamentary dust extinction within APOGEE fields. We also imply a selection on the orbital eccentricity of stars, allowing us to study the contribution to the halo by populations which are likely to be of an \emph{ex-situ} origin. Our method closely follows that outlined in \citet{2016ApJ...818..130B,2016ApJ...823...30B} and \citet{2017arXiv170600018M}, however, we make some key updates to account for changes in the APOGEE target selection in DR14 (and APOGEE-2 in general, as described further in Appendix \ref{sec:appA}). We also show here that correcting for the selection effects due to our kinematic cuts is not necessary, and describe our rationale for this. 

As before, we model APOGEE star counts as a Poisson point-process, considering stars to be distributed in the space defined by $O = [l,b,D,\mu_{l,b},v_{\mathrm{LOS}},H,[J-K_S]_0,\mathrm{[Fe/H]},\mathrm{[Mg/Fe]}]$ with an expected rate $\lambda(O|\theta)$, where $\theta$ is the vector of parameters which are to be determined, defined by a given model of the rate function. The rate function is expressed fully in this case as
\begin{multline}
\label{eq:rate}
    \lambda(O|\theta) = \nu_*(X,Y,Z|\theta) \times f(\vec{v}|\theta) \times|J(X,Y,Z,\vec{v};l,b,D,\mu_{[l,b]},v_\mathrm{LOS})| \\
    \times \rho(H,[J-K_S]_0,\mathrm{[Fe/H]}|X,Y,Z) \times S(l,b,H,[J-K_s]_0,\mu_{[l,b]},v_\mathrm{LOS}) 
\end{multline}
where $\nu_*(X,Y,Z|\theta)$ is the stellar number density in rectangular coordinates, $f(\vec{v}|\theta)$ is the velocity distribution function (DF, which we discuss below),  $\rho(H,[J-K_S]_0,\mathrm{[Fe/H]}|X,Y,Z)$ is the apparent density of stars in colour, apparent magnitude and metallicity space given an $(X,Y,Z)$ position. $|J(X,Y,Z,\vec{v};l,b,D,\mu_{[l,b]},v_\mathrm{LOS})|$ is the Jacobian of the conversion between observed coordinates and Galactocentric cartesian coordinates, which we can split into a purely spatial and a kinematic part as $|J_1(X,Y,Z;l,b,D)\,J_2(\vec{v};\mu_{[l,b]},v_los|l,b,D)|$. $S(l,b,H,[J-K_s]_0,\mu_{[l,b]},v_\mathrm{LOS})$ defines the selection function, which in this case includes a dependence on dereddened $(J-K_s)_0$ colour (due to the colour binning adopted for targeting in DR14), and the orbital eccentricity, through the inclusion of our implied selections in orbital eccentricity. For simplicity, we will factor out the velocity dependency, assuming that $S(l,b,H,[J-K_s]_0,\mu_{[l,b]},v_\mathrm{LOS}) = S_1(l,b,H,[J-K_s]_0) S_2(\mu_{[l,b]},v_\mathrm{LOS})$, as we later show that $S_2(\mu_{[l,b]},v_\mathrm{LOS})$ is likely to be near to unity in specific cases, when only making cuts in, for example, $e$ space. The APOGEE spatial selection function $S_1$ is then a simple 2D piece-wise constant function across magnitude and color bins. We describe the construction of  $S_1(l,b,H,[J-K_s]_0)$ in Appendix \ref{sec:appA}.

Following from the expression of the rate function, and using the assumption that the rate $\lambda(O|\theta)$ only depends on $\theta$ through $\nu_*(X,Y,Z|\theta)$ (as we assume a constant velocity distribution, discussed later), we can then express the log-likelihood of the Poisson point process as 
\begin{align}
    \label{eq:likelihood}
    \ln \mathcal{L}(\theta|O_i) = \sum_{i} \left[ \ln v_*(X_i,Y_i,Z_i|\theta) - \ln \int dO \lambda(O|\theta) \right].
\end{align}
The integral on the right hand side describes the effective observable volume of the survey, and is, for APOGEE, a sum of integrals over each field in the survey. For this application, the effective volume is expressed as
\begin{multline}
    \label{eq:effvol}
    \int dO \lambda(O|\theta) = \sum_{\mathrm{fields}} \Omega_f \int dD D^2\nu_*([X,Y,Z](D,\mathrm{field})|\theta) \\
    \times \mathfrak{S}(\mathrm{field},D) \times \mathfrak{S}_e(\mathrm{field},D),
\end{multline}
where $\nu_*([X,Y,Z](D,\mathrm{field})$ is the density, as before, but evaluated along the line of sight of an APOGEE field. $\mathfrak{S}(\mathrm{field},D)$ and $\mathfrak{S}_e(\mathrm{field},D)$ are the spatial and kinematic 'effective' selection functions. The spatial effective selection function is given by the integration of the raw selection function $S_1$ over the area of the field and the colour magnitude distribution of the tracer, folding in the effect of dust extinction, and is given (for APOGEE DR14) by
\begin{equation}
\begin{split}
    \mathfrak{S}(\mathrm{field,D}) & = \iint dM_{H}\ d(J-K_s)_0\ S_1(\mathrm{field},M_H,(J-K_s)_0)\\
    & \times \rho(H, (J-K_S)_0)\\
    & \times \frac{\Omega_{j,k}(H_{\mathrm{[min,max]},j,k},M_H,A_H[l,b,D],D)}{\Omega_f}.
\end{split}
\end{equation}
The integral under $\rho(H, (J-K_S)_0)$ is performed in the Monte Carlo integration described below. $S_1(\mathrm{field},M_H,(J-K_s)_0)$ is the APOGEE selection function, as previously defined. The third term within the integral is the fractional area of the APOGEE field where stars were observable given the dust extinction $A_H$ and at a given distance modulus and colour.  The observable area $\Omega_{j,k}$ is
\begin{multline}
    \Omega_{j,k}(H_{\mathrm{[min,max]},j,k},M_H,A_H[l,b,D],D) = \\
    \Omega(H_{\mathrm{min},j,k}-[M_H-\mu(D)] < A_H(l,b,D) < H_{\mathrm{max},j,k} - [M_H-\mu(D)])
\end{multline}
where $H_{\mathrm{[min,max]},j,k}$ are the $H$ magnitude limits for a colour-magnitude bin $j,k$ and $\mu(D)$ is the distance modulus at $D$, defined in the usual way. $A_H(l,b,D)$ is taken from a 3D dust map from \citep{2016ApJ...818..130B}, implemented in the \texttt{mwdust} python package\footnote{\url{https://github.com/jobovy/mwdust}} and based on a combination of dust maps provided by \citet{2003A&A...409..205D}, \citet{2015ApJ...810...25G} and \citet{2006A&A...453..635M}. The integral is evaluated by Monte Carlo sampling the underlying colour-magnitude distribution of the giant star tracers which we generate using the PARSEC stellar evolution models \citep{2012MNRAS.427..127B,2017ApJ...835...77M}, weighted by a \citet{2001MNRAS.322..231K} IMF.

The kinematic effective selection function $\mathfrak{S}_e(\mathrm{field},D)$ is given by
\begin{equation}
\begin{split}
    \mathfrak{S}_e(\mathrm{field},D) & = \int d\vec{v} f(\vec{v}) \times |J_2(\vec{v};\mu_{[l,b]},v_los|l,b,D)|\\
    & \times S_2(e|[X,Y,Z](\mathrm{field},D),\mu_{l,b}, v_{\mathrm{LOS}}) .
\end{split}
\end{equation} Ideally, one would fit the entire distribution function $f(\vec{x}, \vec{v})$ to the data, and would then compute $\mathfrak{S}_e$ on the fly, for a given velocity DF $f(\vec{v})$. Performing this integral is computationally expensive, and can be avoided with a careful selection in eccentricity which exploits the fact that the high $e$ halo stars are strongly distinct in their velocity distribution functions. If the velocity distributions of the populations defined by cuts in eccentricity are sufficiently distinct, then $S_2 \sim 1$ over the range where $f(\vec{v})$ is non-zero, and hence $\mathfrak{S}_e \sim 1$. Therefore, we can drop the eccentricity selection correction if the velocity distributions of the populations measured are distinct. We find that for an eccentricity cut at $e = 0.7$ this assumption is valid, as the velocity distributions of stars above and below this cut are quite distinct \citep[this is examined, for example, in][]{2018arXiv180800968M}. 


We optimise the likelihood function (Equation \ref{eq:likelihood}) for MAPs at different eccentricity using a downhill-simplex algorithm. We use the optimal set of parameters $\theta$ to initialise a Markov Chain Monte Carlo (MCMC) sampling of the posterior PDF using the affine-invariant ensemble sampler implemented in the python package \texttt{emcee}  \citep{goodmanweare2010,2013PASP..125..306F}. In the following text, we refer to the $16^{\mathrm{th}}$, $50^{\mathrm{th}}$ and $84^{\mathrm{th}}$ percentile of these samples for the best fit parameters and their uncertainty. We find the uncertainty on values derived from these parameters, such as the mass, by randomly selecting parameter values from these posterior samples and then taking the percentiles of the derived values.


\subsection{Density models}
\label{sec:densitymodels}
Without any good prior knowledge for the density profile of MAPs in the Galactic halo, we aim to be as general as possible when selecting a model for the density. Recent work modelling the Galactic halo using RR Lyrae found that the halo is well modelled by a triaxial density ellipsoid that has a mild elongation along the major axis, which is angled at roughly $70^{\circ}$ with the axis that connects the sun and the Galactic center \citep{2018MNRAS.474.2142I,2019MNRAS.482.3868I}. We adopt for each MAP a simple, single power law (SPL) profile, which is also the best fitting density model identified by \citet{2018MNRAS.474.2142I}. To account for our ignorance regarding the extent of the sample outside of the range observed, we include an exponential `cut-off' term in our density model, which has a scale parameter $\beta$ (corresponding to a scale-length $h_{r_e} = 1/\beta$). We also include here an additional exponential disk term in the density, to account for any contamination from the thicker components of the high \afe{} disk. We use a single exponential disc of density $\nu_{*,\mathrm{disc}}$ with scale height $h_z = 0.8$ kpc and scale length $h_R = 2.2$ kpc \citep{2017arXiv170600018M}, which contributes a fraction $f_\mathrm{disc}$ to the amplitude at the solar position. We later find that $f_\mathrm{disc} \sim 0.1$ to $0.2$, and so an important fraction of the density (and therefore, the stellar mass) in at the high end of the \feh{} range is explained by such a disc model. This profile has the functional form
\begin{equation}
    \nu_*(r_e) \propto (1-f_\mathrm{disc})r_e^{-\alpha}\exp(-\beta r_e)+f_\mathrm{disc}\nu_{*,\mathrm{disc}}. 
\end{equation} In this case we normalise the density such that $\nu_*(r_{e,0})$, the density at the sun, is always unity. The coordinate $r_e$ defines the ellipsoidal surfaces on which the density is constant, themselves defined by
\begin{equation}
    r_e^2 = X_g^2 + \frac{Y_g^2}{p^2} + \frac{Z_g^2}{q^2}
\end{equation} such that $p$ and $q$ describe the $Y_g$-to-$X_g$ and $Z_g$-to-$X_g$ axis ratios, respectively, where $(X_g,Y_g,Z_g)$ are cartesian coordinates relative to the Galactic center. \citet{2018MNRAS.474.2142I}  found that their best fit model also included an allowance for the variation of $q$ with $r_e$ with a scale length of $\sim 12$ kpc. We choose to assume that $q$ is constant as a function of $r_e$ in a given MAP, to facilitate an assessment the variation in flattening as a function of element abundances and kinematics. The shape parameters $q$ and $p$ are both constrained to be $< 1$, forcing the longest axis to be that in the $X_g$ direction. We allow the orientation of the density ellipsoid to vary, applying a transformation defined by the unit vector $\hat{z}$ along the transformed $z$ axis, and the angle of rotation (from the original $x$ axis) of the ellipsoid about this transformed axis, $\phi$. In practice, we ensure the $z$-axis unit vector is sampled uniformly by de-projecting samples from an equal-area rectangular projection of the unit sphere. Using this method, the unit vector is represented by two parameters, $\eta$ and $\theta$, such that $\hat{z} \;=\; \left(\sqrt{1-\eta^2}\cos \theta,\; \sqrt{1-\eta^2}\sin \theta,\; \eta\right)$, where $\eta$ is sampled uniformly between -1 and 1, and $\theta$ between 0 and $2\pi$. As an example, $\eta=1, \theta=0$ and $\phi = 0$ describes perfect alignment with the Galactocentric cartesian system. This transformation generally has little impact on the measurement of the total mass in a MAP, as the parameters tend to define an ellipsoid which is in rough alignment with the solar position. We adopt uninformative priors on all the parameters, and only set the allowed range of $\alpha$ to be positive, and the rest of the parameters to be in the range $[0,1]$.

\begin{figure*}
\includegraphics[width=\textwidth]{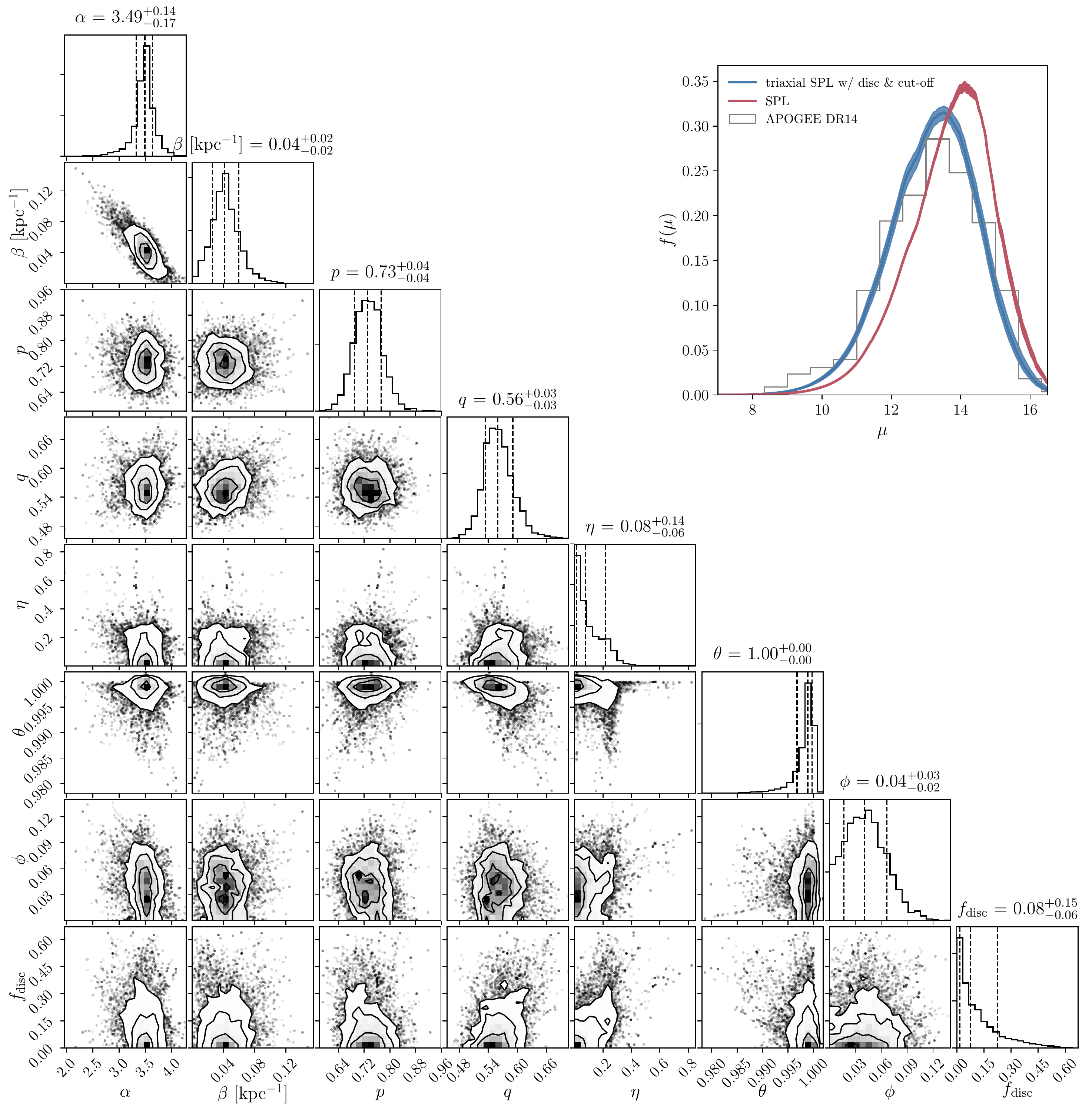}
\caption{\label{fig:full_corner} Corner plot showing posterior samples of the parameters for the adopted triaxial single power law (SPL) model when fit to the full statistical sample between $-3 <\mathrm{[Fe/H]} < -1$, and $0.0 < \mathrm{[Mg/Fe]} < 0.4$. The posterior distributions are well behaved. The best fit model has a steep power law $\alpha \sim 4$, and is slightly flattened along its $y$ and $z$ axes, i.e. $q$ and $p$ are less than 1. The inset panel (top right) shows the distance modulus $\mu$ distribution predicted by this model and a simple spherical power law with the same $\alpha$, compared to the APOGEE data, demonstrating the better quality fit afforded by adopting a triaxial density profile. }
\end{figure*}

\subsection{Mass Measurement}
\label{sec:mass}
Once a well-fitting density model and its uncertainty is found for a MAP, we measure its mass by computing the normalisation of the rate function in Equation (\ref{eq:rate}). We calculate the expected number of stars seen by APOGEE for a given density model normalised to 1 at the solar position, $N(\nu_{*,0}=1)$, by integrating the rate function over the observable volume of the survey. In practice, this integral takes the form
\begin{multline}
    N(\nu_{*,0} = 1) =
    \int_{\mathrm{fields}} d\mathrm{field}\ dD \lambda(\mathrm{field}, D) = \\ \int d\mathrm{field}\ d\mu\ \frac{D^{3}(\mu)\log(10)}{5}\nu_*([R,\phi,z](\mathrm{field},\mu)|\theta)\times \mathfrak{S}(\mathrm{field},\mu), 
\end{multline} as we compute the density and effective selection in APOGEE sightlines on a grid linearly spaced in distance modulus $\mu$. Since the true number of observed stars $N_\mathrm{obs} = A\ N(\nu_{*,0} = 1)$, comparison of the expected number count for a normalised density model with the true observed number of stars in a given MAP provides the proper amplitude, $A$, which is then equivalent to the true number density of giants at the sun, $\nu_{*,0}$, which depends on the MAP considered.

We then convert this number density to a stellar mass-density using stellar evolution models, which facilitate the conversion of number counts in giant stars to the mass of the entire underlying population. We use the PARSEC isochrones \citep{2012MNRAS.427..127B,2017ApJ...835...77M}, weighted with a \citet{2001MNRAS.322..231K} IMF. We calculate the average mass of giants $\langle M_{\mathrm{giants}} \rangle$ observed in APOGEE by applying the equivalent of the observational cuts (from the APOGEE targeting procedure) in $\mathrm{(J-K_S)_0}$ and our imposed cut between $1 < \log(g) < 3$ to the isochrones and finding the IMF weighted mean mass of the remaining points. Similarly, we find the fraction of stellar mass in giants $\omega$ by taking the ratio between the IMF weighted sum of isochrone points within these boundaries and those outside. The conversion factor between giant number counts and total stellar mass for each MAP is then simply calculated as 
\begin{equation}
    \chi(\mathrm{[Fe/H]}) = \frac{\langle M_{\mathrm{giants}} \rangle(\mathrm{[Fe/H]})}{\omega(\mathrm{[Fe/H]}}.
\end{equation}
We compute this factor for each field and each selection in \feh{}, adjusting the limit in $(J-K_S)_0$ to reflect the minimum $(J-K_S)_0$ of the bluest bin adopted in that field, and only considering isochrones which fell in the limits of \feh{} for each MAP. The edges in colour binning for each field are then accounted for by our integration under $\rho((J-K_S)_0,H)$ for the effective selection function. This factor can be as large as $\sim600\ \mathrm{M_{\odot}\ \mathrm{star}^{-1}}$ for the lowest \feh{} bins, in fields where the APOGEE $(J-K_S)_0$ limit was $0.5$. For higher metallicity bins, and in fields where APOGEE assumed a bluer $(J-K_S)_0 > 0.3$ cut, this number approaches $\sim200\ \mathrm{M_{\odot}\ \mathrm{star}^{-1}}$. We note that this method does not account for populations which are not represented in the adopted stellar models. One example of such a population are AGB-\emph{manqu\'e} stars \citep[e.g.][]{2010A&A...522A..77G}, which are expected to be present at lower \feh{}. We expect these effects to be smaller than our statistical uncertainties, however, and demonstrate in Appendix \ref{sec:hst}, using Hubble Space Telescope photometry that the factors which we determine here are reliable against any systematic uncertainty arising from the stellar evolution models.
Combining these factors with the number density normalisation, we attain the appropriate halo mass normalisation, $\rho_0 = \chi(\mathrm{[Fe/H]})\  \nu_{*,0}(\mathrm{[Fe/H], [Mg/Fe]}, e)$, for each MAP.

Once we gain the normalisation for a given MAP, we integrate the (now properly normalised) density models described by 400 samples from the posterior distributions of their parameters to attain the total mass within a population. To avoid over-extrapolating from our fits to the halo density, we only integrate for the mass between $2<r<70$ kpc. Summing MAPs together provides an estimate of the total halo mass within 70 kpc, in the range of \feh{}, \mgfe{} and $e$ considered.

\section{Results}
\label{sec:results}
\subsection{Initial fit to the full halo sample}
We first perform the fitting procedure described in Section \ref{sec:method} on broadly defined, larger samples of likely halo stars in APOGEE, to ensure that the density profile we define provides a good fit to the data, and facilitate some initial conclusions to be drawn. This test also provides insights into the conclusions which would be drawn if we did not have the ability to divide the stellar halo into its constituent mono-abundance populations. We select all stars between $-3 < $ \feh{} $< -1$ and $0 < $ \mgfe{} $ < 0.4$, and perform the single power law fit to the data. The resulting best fit profile has a steep power law, with $\alpha=3.5\pm 0.2$ which has a cut-off scale length of $h_{r_e} = 20\pm 10\ \mathrm{kpc}$ and is slightly flattened along its $y$ and $z$ axis, with $p = 0.73 \pm 0.04$ and $q = 0.56 \pm 0.03$. The $z$-axis tilt (rotation around the $x$ or $y$ axes) is very minor, but we find a very slight rotation about the $z$-axis with $\phi =  7\pm 3$ deg. A corner plot showing samples from the posterior distribution of parameters given the data is shown in Figure \ref{fig:full_corner}, and demonstrates that the posterior distributions are well behaved. The inset panel in that figure (\emph{top right}) shows the distance modulus $\mu$ distribution predicted by the best fit model in blue, compared to that from a spherical power law model with the same $\alpha$, and compared to the actual APOGEE data, and demonstrates the better fit afforded by adopting the triaxial model.

From this profile, using the APOGEE star counts, we gain a halo normalisation $\rho_0 = 6.3^{+0.5}_{-0.4}\times10^{-5}\ \mathrm{M_\odot\ pc^{3}}$. The total mass which we find for this profile, following the procedure outlined in Section \ref{sec:mass}, is $M_{\mathrm{tot}} = 4.8^{+0.5}_{-0.6}\times10^{8}\ \mathrm{M_\odot}$. The total mass we obtain from this exercise is broadly consistent with that provided in \citet{2016ARA&A..54..529B}, based on MSTO counts by \citet{2008ApJ...680..295B}. It is interesting to compare the best fit scale length of our exponential cutoff to the break radius which is commonly discussed in the literature, at around 20 to 30 kpc \citep[e.g.][]{2011MNRAS.416.2903D,2011ApJ...731....4S,2009MNRAS.398.1757W}. We tested a fit which replaced the exponential cut-off with a break in the power law profile, such that outside a break radius $r_{e,b}$ the density becomes proportional to $r_e^{\alpha_{\mathrm{out}}-\alpha_{\mathrm{in}}}r_e^{-\alpha_\mathrm{out}}$. The best fit profile has $\alpha_\mathrm{in} = 3.9 \pm 0.1$, with a break at $r_{e,b} = 21\pm 1$ kpc, outside which the density drops steeply with $\alpha_{\mathrm{out}}=5.5\pm 0.4$. However, the single power law with an exponential cut-off provides a marginally better fit to the data, with a higher maximum likelihood. Given that the single power law profile more simply represents the data, we choose to move forward to fit this model to the MAPs.

\begin{figure*}
\includegraphics[width=\textwidth]{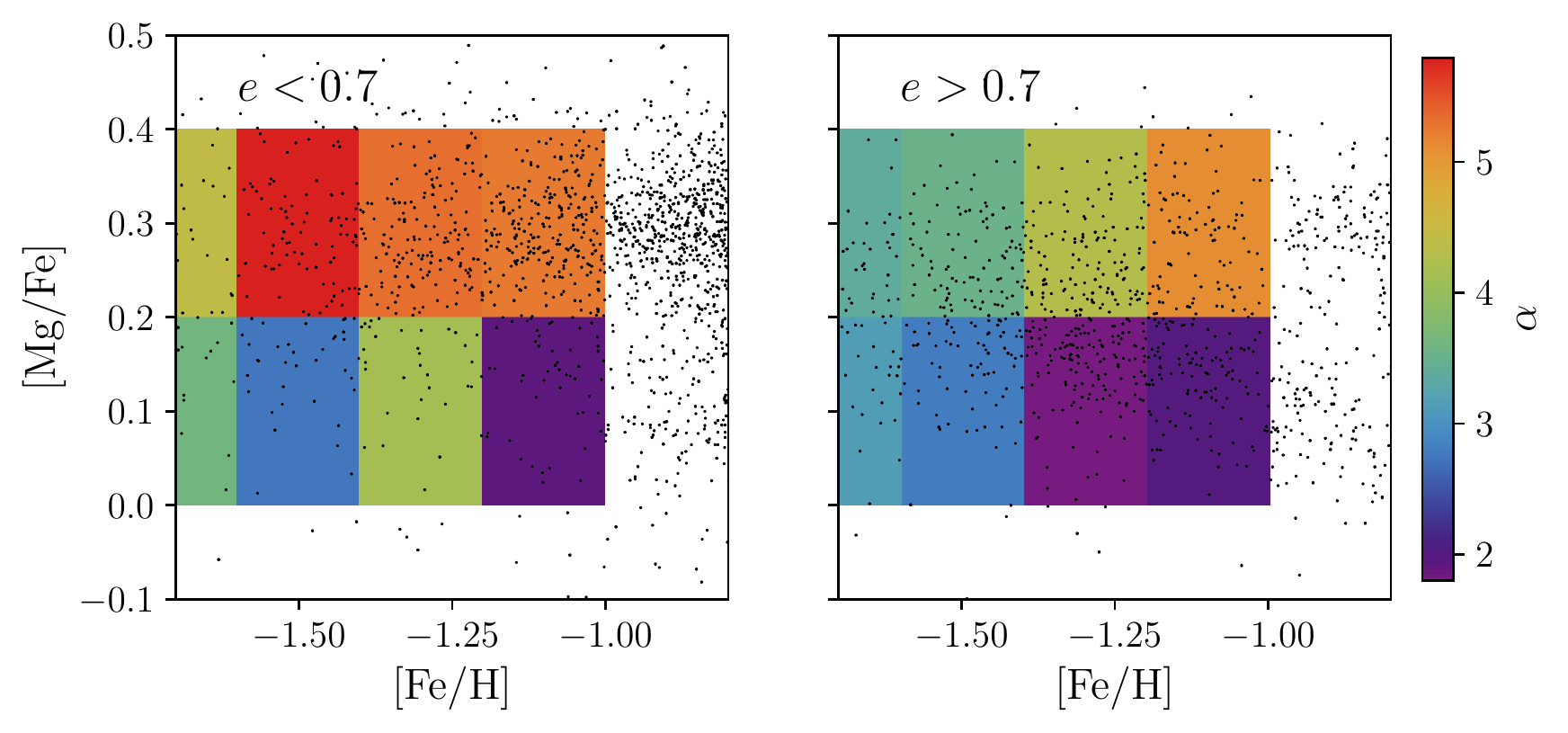}
\caption{\label{fig:alpha_MAPs} The power law index $\alpha$ of MAPs in the high ($e > 0.7$) and low ($e < 0.7$) eccentricity stellar halo. At both high and low $e$, the halo cleanly splits into a high and low \mgfe{} component with steeper and shallower (respectively) power laws. At the lowest \feh{} the power law indices of the high and low \mgfe{} populations converge.}
\end{figure*}

\subsection{Fitting MAPs in the APOGEE halo sample}
\label{sec:MAPs}
We now turn to performing our analysis on chemically defined MAPs. We choose to also divide MAPs based on their kinematics in order to study the mass contributed by stars on near-radial orbits in the halo which, it has been recently suggested, may be the debris of accreted dwarf galaxies and make up a major fraction of the halo stellar mass. In this way, we aim to directly place a constraint on the stellar mass of accreted debris in the halo, be it from a single satellite or multiple.  

We fit the triaxial SPL model to the MAPs as shown in Figure \ref{fig:afe_feh_sample}, dividing each into high ($e > 0.7$) and low ($e < 0.7$) orbital eccentricity bins. We find that a remarkably good fit to the data is returned even for bins with very low $N$, with lower $N$ bins returning a higher uncertainty. All bins have over 15 stars, and trace a range in Galactocentric radius roughly between our integration limit of $2 < r < 70$ kpc. The best fit models in each MAP generally have $f_\mathrm{disc}$ in the region $0.0 < f_\mathrm{disc} < 0.2$, with the only exception being the MAP at low $e$ at the highest \feh{} and highest \mgfe{}, which has $f_\mathrm{disc} = 0.5^{+0.3}_{-0.3}$.

The best fitting power law indices for each MAP at low and high $e$ are shown in Figure \ref{fig:alpha_MAPs}, giving an indication of the halo structure as a function of element abundances and kinematics. It is immediately clear that the high and low \mgfe{} populations in the halo also define different structures, such that at both high and low $e$, the lower \mgfe{} stars tend to have shallow power laws of order $\sim 2$ to $3$, whereas the higher \mgfe{} populations tend to have steeper power laws with $\alpha$ as high as $\sim 5$. The median uncertainty on our determination of $\alpha$ is $\langle \delta \alpha \rangle \sim 0.4$. This clear division in halo structure as a function of \mgfe{} is interesting given that the power indices of the low and high \mgfe{} MAPs are respectively consistent with the inner and outer power law slopes which are commonly found in the literature \citep[e.g.][]{2011MNRAS.416.2903D,2011ApJ...731....4S,2009MNRAS.398.1757W}. It is interesting also that the power law indices appear to converge to $\sim 3 - 4$ at the lowest \feh{} in both the low and high $e$ samples - but we note that these bins consist of stars in a much larger range of \feh{}, from $-3 < \mathrm{[Fe/H]} < -1.6$. We discuss the relationship between this result and previous findings in the stellar halo in Section \ref{sec:break}.

The best fit scale length of the exponential cut-off is remarkably similar between all the MAPs, but is not well constrained in general, owing to the limited extent of the data in each MAP. Combining all samples from each MAP, we find $h_{r_e,\mathrm{tot}} = 26^{+85}_{-15}$ kpc. This level of uncertainty is consistent accross all the MAPs. Comparing the exponential cut off between the high and low $e$ populations, we do not detect any significant differences between the populations. Similarly, we do not find any significant difference in the best fit cut-off scale between the high and low \mgfe{} MAPs. This is interesting given that the drop in the stellar halo density is generally attributed to a shift in the stellar population \citep[e.g.][]{2011MNRAS.416.2903D}. We verify that the exponential cut-off is not simply reflecting the limit of our sample in $r$, finding that there is no correlation between the maximum $r$ in a given MAP, and its $h_{r_e}$. 

\begin{figure}
\includegraphics[width=\columnwidth]{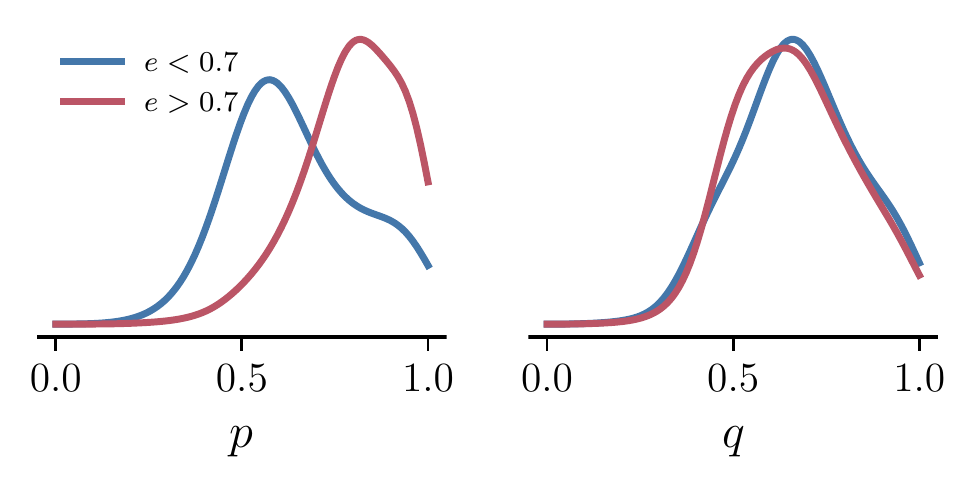}
\caption{\label{fig:triaxiality} Kernel Density Estimates (KDE) of the posterior distribution of the $p$ and $q$ parameters for the combined MAPs in the low and high $e$ samples. The high $e$ MAPs have $p$ closer to 1 (near-axisymmetry) than the low $e$ groups. We find that both eccentricity selected groups are flattened, with similar $q \sim 0.7$. }
\end{figure}

We detect a degree of triaxiality in all the MAPs, with $q$ and $p$ less than 1 in most cases. This is unsurprising, given that the best fit model for the full sample was also triaxial. Similarly to the above, we find that the best fit models are triaxial, but our constraints on the orientation of this ellipsoid are not strong, and can take on orientation anywhere between $0. < \phi < 60$ deg. The angle between the Galactic pole and the ellipsoid's $z$ axes can be anywhere in the region from 0 to 70 degrees. However, the median orientation tends to be such that the ellipsoids are aligned with the solar position, as in the results for the full sample, but we emphasise that this alignment is very poorly constrained by the data. We note that this finding is somewhat suspicious, and consider that it may be due to sample selection effects which cannot be corrected for by our procedure.

Regardless of the orientation, it is interesting to study how the triaxiality changes between the high and low $e$ samples. We show the posterior distribution of $p$ and $q$ for the combined MAPs in the high and low $e$ samples in Figure \ref{fig:triaxiality}.  Low $e$ MAPs have a median $p = 0.6$, whereas the high $e$ population has a far more axisymmetric structure, with median $p = 0.8$. The situation is somewhat different in terms of the flattening, $q$, with the high and low $e$ population assuming similarly flattened structures with median $q = 0.7$. That the high $e$ population takes on a more axisymmetric structure whereas the low $e$ population appears quite strongly triaxial is interesting given the expected origin of the high $e$ population in an accreted satellite.

\begin{figure*}
\includegraphics[width=\textwidth]{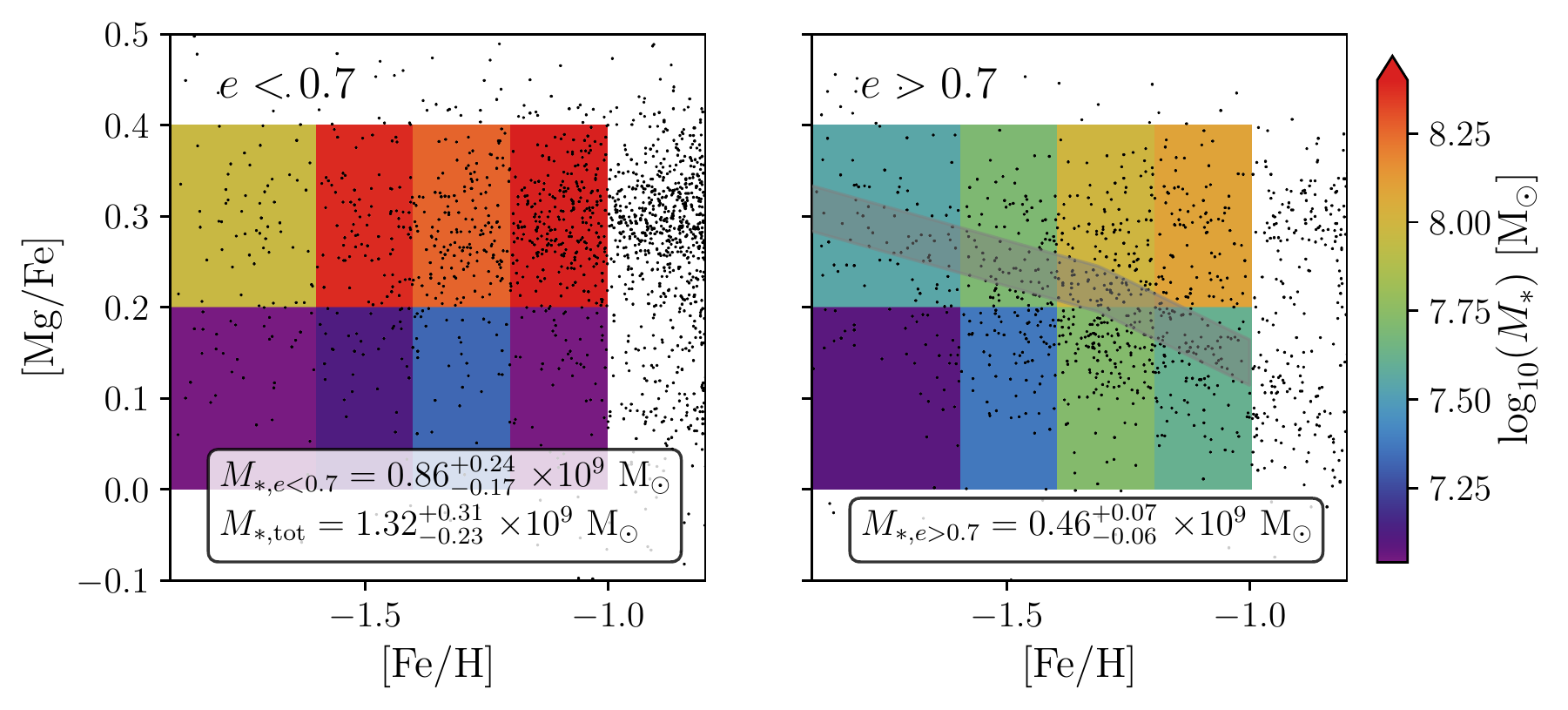}
\caption{\label{fig:mass_MAPs} The mass contribution from MAPs in the high ($e > 0.7$) and low ($e < 0.7$) eccentricity stellar halo. The coloured boxes give the bounds of and mass in each MAP, and the overplotted scatter points show the actual \mgfe{}-\feh{} of the data. In the low $e$ population, much of the mass is concentrated at high \mgfe{} and \feh{}. At high $e$, the mass is spread between low and high \mgfe{} and forms a 'track' which roughly follows that seen in the black points and the \feh{}-\mgfe{} track of the \emph{Gaia-Enceladus}/Sausage population as fit to the DR14 data in \citep{2018arXiv180800968M}, shown here by the gray band. The mass in MAPs which lie on this track provide an estimate of the stellar mass of this population of $M_{*,\mathrm{sausage}} = 3\pm1\times10^{8}\ \mathrm{M_{\odot}}$}
\end{figure*}

Finally, we use the MAPs to constrain the normalisation of the stellar halo density, as well as its total mass between $2 < r < 70$ kpc. Combining all the MAPs at high and low $e$, and including those at the lowest \feh{}, we obtain a total halo normalisation of $\rho_0 = 6.3^{+1.7}_{-1.4}\times10^{-5}\ \mathrm{M_{\odot}\ pc^{3}}$, and a total stellar mass accross our range of \feh{} and \mgfe{} of $M_{*,\mathrm{tot}} = 1.3^{+0.3}_{-0.2}\times10^{9}\ \mathrm{M_{\odot}}$. The stellar mass contributed by each MAP in the low and high $e$ samples is shown in Figure \ref{fig:mass_MAPs}. It is immediately clear that the low $e$ stars contribute the majority of the stellar mass in the halo. The MAPs at high \mgfe{} with low $e$ contribute $7^{+3}_{-3}\times10^{8}\ \mathrm{M_{\odot}}$, which is $\sim 50\%$ of the total stellar mass. The entire high $e$ population contributes only $5\pm1\times10^{8}\ \mathrm{M_{\odot}}$, equivalent to $\sim 35\%$ of the total stellar halo mass. We return to possible systematics on this estimate in Section \ref{sec:lowmass}. It is likely that the high \mgfe{}, high $e$ populations define those at the interface between the high \afe{} disc and the halo, and we find that these populations indeed have the highest $f_\mathrm{disc}$, but we stress that this mass estimate is that of only the component described by the SPL density profile. The estimated mass in disc contaminants is very low across the range of \feh and \mgfe\ considered, of order $10^{7}\ \mathrm{M_\odot}$.

Returning to the high $e$ population, we note that the mass in this population is spread across the range in \mgfe{}, whereas in the low $e$ MAPs, the mass is concentrated at high \mgfe{}. It is noteworthy that the highest mass MAP is still at the highest \feh{} and high \mgfe{}, with $M_{*} = 1.2\pm0.4\times10^{8}\ \mathrm{M_{\odot}}$. Since it seems unlikely that dwarf galaxies might have such abundances, it seems logical to assume that at least some of this mass is of an \emph{in-situ} origin. A number of recent works based on \emph{Gaia} DR2 data have suggested that early disc stars may have been heated into a halo-like configuration and onto more radial orbits by ancient and violent mergers \citep[e.g.][]{2018arXiv181208232D,2019arXiv190904679B}, and this is expected from a theoretical standpoint \citep[e.g.][]{2011MNRAS.416.2802F,2012MNRAS.420.2245M}. Regardless of this excess mass at high \mgfe{} and \feh{}, the stellar mass in the high $e$ population is spread more evenly accross the MAPs at high and low \mgfe{}. This is consistent with the idea that this population is likely to be dominated by stars which were formed in dwarf galaxies which subsequently accreted onto the Milky Way, as it is unlikely that stars formed \emph{in-situ} can achieve \mgfe{} as low as $< 0.2$ at \feh$< -1.$, which is characteristic of the long star formation timescales in such dwarfs. Furthermore, it is interesting to note that the mass distribution of the high $e$ MAPs seems to follow somewhat the \feh{}-\mgfe{} `track' which is visible in the stellar distribution in the underlying black points, if the high \mgfe{}, high \feh{} MAP is excluded. It is likely that these stars belong to the progenitor of the \emph{Gaia-Enceladus}/Sausage population \citep{2018MNRAS.478..611B,2018arXiv180606038H}, therefore, the combined mass of these MAPs gives an estimate of the stellar mass of that progenitor. Summing the mass in each MAP which the \emph{Gaia-Enceladus}/Sausage \feh{}-\mgfe{} track \citep[based on modelling in][]{2018arXiv180800968M} goes through, we estimate this mass to be $M_{*,\mathrm{sausage}} = 3\pm1\times10^{8}\ \mathrm{M_{\odot}}$.

\section{Discussion}
\label{sec:discussion}
\subsection{Reconciling broken and single power laws with MAPs}
\label{sec:break}
An exhaustive comparison of our measurements with those from the literature is made difficult by the quite different way in which we consider the halo here, but we will attempt here to make links with more recent works on the stellar halo. In particular we attempt to show how our analysis of MAPs offers insights into the underlying halo structure, reconciling the varying measurements found in the literature. 

Most earlier works fit single power law models to samples of tracer stars, using, for example, Main Sequence Turn Off (MSTO) stars \citep[e.g.][]{2000AJ....119.2254M} or Blue Horizontal Branch (BHB) stars \citep[e.g.][]{2000ApJ...540..825Y}, the photometry of which is relatively easily calibrated to provide accurate distances. Such works tend to require some flattening (e.g. $q\sim 0.6$) to fit star counts well, and agree on an exponent of $\alpha \sim 3$. A seminal work by \citet{2011MNRAS.416.2903D} established that BHB stars in SDSS were well fit by a broken power law, such that the profile slope is shallow, $\alpha_{\mathrm{in}} \sim 2.3$, out to a break radius of $r_b = 27$ kpc, after which the profile steepens, such that $\alpha_{\mathrm{out}} \sim 4.6$. However, a later study, using SDSS-SEGUE giants showed that such a profile can be reconciled with a single power law, but whose flattening parameter $q$ varied with radius between $q \sim 0.5$ and $0.8$ \citep{2015ApJ...809..144X}. The overall slope of their best fit variable flattening profile is rather steep, with $\alpha = 4.2$. A more recent look at RR Lyrae in the inner halo found that a similar profile provided a good fit to these data, such that $\alpha = 2.96$ with $0.57 < q < 0.75$ - although this study concentrated on the innermost parts of the halo \citep{2018MNRAS.474.2142I}. 

Similar behaviour is also evident in the profiles we fit to the APOGEE data. Fitting the entire sample, we find that the halo density is best fit such that it has an exponential cut off with a scale $h_{r_e} = 20\pm 10$ kpc. Fitting individual MAPs, we showed that the shape parameter $p$ is dependent on the eccentricity and element abundances of the populations, such that the lower $e$, high \mgfe{} populations are more triaxial and have a steeper power law slope, whereas the high $e$ populations have shallow power laws and are more axisymmetric, with $p \sim 0.8$ and $q \sim 0.5$. Combining all the MAPs, at all eccentricities, we find a that the total power law slope is consistent with $\alpha \sim 3 - 4$, in consistency with our fit to the full sample. The cut-off radii of the MAPs is not determined at high significance, but the combined value is consistent with that found using the full sample. Thus, we can establish that the inner and outer profile commonly discussed in the literature is most likely to be a manifestation of the combined spatial structure of its constituent stellar populations, which themselves are likely to arise from both \emph{in-situ} and accreted populations. 

We also note that our measurement of the halo normalisation $\rho_0$ is consistent between the fit to the full sample and that obtained from the combination of all the MAPs, yet we find significantly different estimates of the total stellar mass in the halo between these fits, obtaining $M_{\mathrm{tot}} = 4.8^{+0.5}_{-0.6}\times10^{8}\ \mathrm{M_\odot}$ for the entire sample, and $M_{*,\mathrm{tot}} = 1.3^{+0.3}_{-0.2}\times10^{9}\ \mathrm{M_{\odot}}$ for the combined MAPs -- almost a factor of 10 greater. Previous estimates, based on fitting density models to broad samples of halo stars, have placed the stellar mass in the halo between 4 and $7\times10^{8}\ \mathrm{M_\odot}$ \citep[see, e.g.][and references therein]{2016ARA&A..54..529B}, in consistency with our mass estimate based on a single profile fit. A recent paper by \citet{2019arXiv190802763D} estimated a mass for the halo $M_* = 1.2 \pm 0.3 \times10^9\ \mathrm{M_\odot}$, based on a normalisation of $6.9\times10^{-5}\ \mathrm{M_\odot pc^{-3}}$ (we compare with their fit excluding the Sagittarius dwarf, as it is likely that our estimate also excludes this population). Adopting our value of the normalisation $\rho_0 = 6.3^{+1.7}_{-1.4}\times10^{-5}\ \mathrm{M_{\odot}\ pc^{3}}$, we yield a total stellar mass in the halo using the \citet{1965TrAlm...5...87E} profile adopted by \citet{2019arXiv190802763D}, of $1.0^{+0.3}_{-0.2}\times10^{9}\ \mathrm{M_\odot}$, consistent with our estimate based on the combined MAPs. 


\subsection{Assessing the mass of accreted stars with $e > 0.7$}
\label{sec:lowmass}
By measuring the stellar mass enclosed in MAPs in kinematically defined populations with orbital eccentricity $e$ higher or lower than $0.7$ we have determined that stars with $e > 0.7$ account for only $\sim 35\%$ of the total stellar mass in the Milky Way between $-3 < \mathrm{[Fe/H]} < -1$. At face value, this appears to be in contention with recent results suggesting that the radially biased population makes up at least $50\%$ of the mass within 25 kpc \citep[e.g.][]{2019MNRAS.486..378L}, and that the progenitor of the accretion event which created it was of a relatively high mass, between 0.5 and $2\times10^{9}\ \mathrm{M_{\odot}}$ \citep{2018arXiv180606038H,2019arXiv190309320D,2019MNRAS.487L..47V}. We now discuss the possible systematic uncertainties implied on this fraction by our choice of eccentricity cut, and then discuss the extant literature in light of our results.

The present selection in eccentricity is made on the assumption that $e = 0.7$ provides the best division between the radially biased component of the halo \citep[the \emph{Gaia}-Sausage, e.g.][]{2018MNRAS.478..611B} and the other, more isotropic components. The eccentricity distribution of stars in our sample clearly has a distinct peak concentrated above this value, as shown in Figure \ref{fig:eccdist}. It is conceivable that at least some of the stars in the true high eccentricity population have rather large uncertainties, which might lead them to be counted in the $e< 0.7$ population in our measurements. We already discussed the effects of uncertainty in the astrometric parameters at large distances in Section \ref{sec:data}. We now further test this scenario by re-fitting the MAPs with eccentricity cuts at $e = 0.6$ and at $e = 0.8$, assessing the systematic uncertainty on the fractional contribution to the stellar halo mass by high eccentricity stars. The general results remain unchanged under these selections, but the mass fraction at high and low $e$ changes due to the re-allocation of star counts. We find that the fraction of mass above the eccentricity cut changes near-linearly such that it goes from 50 to 20 between $e > 0.6$ and $e > 0.8$. We can thus only be highly confident that the fraction of the total stellar mass in the halo contributed by the radially biased population is $ < 50\%$, and place a constraint on its contribution to the total stellar halo mass (including this systematic uncertainty) of $f_{\mathrm{high}\ e} = 0.35^{+0.04}_{-0.02}\mathrm{(stat.)}^{+0.15}_{-0.15}\mathrm{(syst.)}$.

We stress that our estimate of the fraction of mass contributed by the radially biased population as discovered by \citet{2018MNRAS.478..611B} is likely to still be an upper limit, as we include here mass which is found lying at high \mgfe{} and high \feh{}. Following from our brief discussion in Section \ref{sec:MAPs} regarding the heating of populations which were present in the Milky Way before any major merging events \citep[e.g. found recently in the \emph{Gaia} DR2 data:][]{2018arXiv181208232D,2019arXiv190904679B}, we discern that stars in this region of \mgfe{}-\feh{} space are likely to be made up -- at least in-part -- by such populations and also likely overlap with the true accreted stars in kinematics space (particularly in $e$). If we make the crude assumption that all the mass in the MAP with $-1.2 < \mathrm{[Fe/H]} < -1.$ and $0.2 < \mathrm{[Mg/Fe]} < 0.4$ belongs to this heated disc population, then we find that the fraction of high $e$ mass contributed by potentially heated stars is $f_{\mathrm{heated}, e > 0.7} = 0.28\pm 0.03\ \mathrm{(stat.)}\ \pm 0.02\ \mathrm{(syst.)}$. While it is unlikely that the fraction contributed by the radially biased population found by \citet[][]{2019MNRAS.486..378L} was contaminated by these heated stars, as that study employed BHB star tracers (which likely do not extend to these high \feh{}), future studies of the accreted populations in the halo which do not have access to spectroscopic constraints will need to take any contribution from these stars into account. We note that our estimate of the total halo stellar mass includes these populations. 

Excluding the fraction of the stellar mass at $e > 0.7$ which we have attributed to heated disc populations, we arrive at a more robust estimate for the mass in accreted stars with $e > 0.7$ -- which are likely to be associated with the \emph{Gaia-Enceladus}/Sausage progenitor -- of $M_{*,\mathrm{accreted}, e > 0.7} = 3\pm 1\ \mathrm{(stat.)} \pm 1\ \mathrm{(syst.)}\times10^{8}\ \mathrm{M_\odot}$. We note that this is consistent with the rough estimate for the \emph{Gaia-Enceladus}/Sausage mass which we presented at the end of Section \ref{sec:results}, and is at the low end of the estimate of the \emph{Gaia-Enceladus}/Sausage progenitor mass found on the basis of element abundances in the EAGLE simulations, in \citet{2018arXiv180800968M}. It has been shown in a number of works that accreted stars at high $e$ are likely part of the same single dwarf \citep{2018arXiv180606038H,2018MNRAS.478..611B,2018arXiv180800968M,2019MNRAS.488.1235M} so we are qualified to make the claim that this is at least an upper limit to the stellar mass of this massive accretion event. It is noteworthy that previous estimates have constrained the stellar mass of this dwarf, using indirect methods, to be rather higher, of the order $10^9\ \mathrm{M_\odot}$ \citep{2018MNRAS.478..611B,2019arXiv190309320D,2018arXiv180606038H,2018arXiv180800968M,2019MNRAS.488.1235M,2019MNRAS.487L..47V}. Using the relation between stellar mass and halo mass derived by \citet{2017MNRAS.467.2019R} \citep[and used to derive an estimate of the stellar mass in][]{2019arXiv190309320D}, we arrive at an estimate of the progenitor halo mass of $M_{200} \sim 10^{10.2\pm 0.2}\ \mathrm{M_\odot}$

Furthermore, having identified the stellar population linked to the \emph{in-situ} halo, we can also place constraints on the fraction of mass in the entire halo (independent of kinematics) which is made up of likely accreted stars. If we make the assumption that the \emph{only} \emph{in-situ} halo stars in the halo lie at the highest \mgfe{} and \feh{} {\red \citep[which is likely robust, given the extant literature on abundances in the halo, e.g.][]{2010A&A...511L..10N,2016MNRAS.458.2541S,2018ApJ...852...50F,2018ApJ...852...49H}}, and thus that the rest of the stars in this range of element abundances were members of accreted dwarfs {\red and are therefore distinct in \mgfe{} and \feh{} \citep[see e.g.][]{2007PASP..119..939G,2016MNRAS.458.2541S}}, then we can compute this fraction. For our fiducial $e = 0.7$ division, we find that the mass outside of the highest \mgfe{}-\feh{} bin is $M_{*,\mathrm{accreted}} = 9^{+2}_{-1}\times 10^{8}\ \mathrm{M_{\odot}}$, which equates to $f_{\mathrm{accreted}} = 0.7\pm0.1$. This indicates that the Milky Way stellar halo is dominated by accreted stars. Of course, this also relies on the assumption that the mass in \emph{in-situ} stars is very low outside the highest \mgfe{}-\feh{} MAP. It is likely that our $M_{*,\mathrm{accreted}}$ is contaminated in part by this mass, but we make the fair assumption that this contaminant mass is likely well within our stated uncertainty.  We note that our $M_{*,\mathrm{accreted}}$ is more consistent with our re-estimation of the \citet{2019arXiv190802763D} mass -- which makes sense, as that study avoided low latitudes where heated \emph{in-situ} stars should be more dominant. 

This estimate also then implies that the mass of the \emph{in-situ} halo at $\feh < -1.0$ is $M_{*,\mathrm{\emph{in-situ}}} = 4\pm2\times10^{8}\ \mathrm{M_{\odot}}$. This mass is a very small fraction of the current Milky Way disc mass, of order $1\%$, implying that this heating likely had quite a minor effect on the overall formation and evolution of the disc. We note however, that this does not account for heated disc mass at higher \feh{}, and defer a further exploration of the heated disc to future studies.

Finally, Our estimate of the total accreted mass allows us to find the contribution of \emph{Gaia-Enceladus}/Sausage progenitor stars to the accreted halo, rather than the total halo. This allows for a better comparison with studies that likely miss the contribution from higher \feh{} stars. Taking the fraction of mass at high $e$ in the absence of the high \mgfe{}, high \feh{},  we find that between 30 to 50 per cent of the accreted stellar halo is at $e > 0.7$, in line with (but at the low end of) the results of \citet{2019MNRAS.486..378L}.

\section{Summary and Conclusions}
\label{sec:conclusions}
In this paper we have assessed the contribution to the stellar halo by stellar populations as defined by their element abundances and kinematics. A summary of our results is as follows:
\begin{itemize}
    \item We determine, using a simple Gaussian Mixture Model, that stars with \feh{}$ < -1.$ are divided cleanly by a cut in orbital eccentricity at $e = 0.7$ (see Figure \ref{fig:eccdist}).
    \item We show that once the survey selection effects are accounted for (see Appendix \ref{sec:appA}), the APOGEE data between $-3 < \mathrm{[Fe/H]} < -1$ are fit well by a triaxial density ellipsoid, defined by a single power law with an exponential cut-off. We find that the best fit power law slope is $\alpha = 3.5\pm 0.2$, with a cut-off scale length of $h_{r_e} = 20\pm 10$ kpc, which is broadly consistent with the commonly determined break radius between 20 and 30 kpc. The best fit model is triaxial with $p \sim 0.7$ and $q \sim 0.6$, with only a mild disalignment from the position of the sun (see Figure \ref{fig:full_corner}).
    \item By fitting individual MAPs with the same density model, we reveal that the structure of the halo depends strongly on \feh{}, \mgfe{} and orbital eccentricity $e$. High \mgfe{} MAPs are well fit by steep power laws $\alpha \sim 4$, whereas the lower \mgfe{} stars have shallower power law slopes with $\alpha \sim 2$ (see Figure \ref{fig:alpha_MAPs}).
    \item The triaxiality of the stellar halo appears to change with eccentricity, with low $e$ MAPs occupying more strongly triaxial density ellipsoids with $p \sim 0.5$ and $q \sim 0.6$, whereas the high $e$ populations have $p \sim 0.8 - 1.$ and are more axisymmetric (see Figure \ref{fig:triaxiality}).
    \item We determine the total stellar mass of the halo between $-3 < \feh{} < -1$ and $0.0 < \mgfe{} < 0.4$ to be $M_{*,\mathrm{tot}} = 1.3^{+0.3}_{-0.2}\times10^{9}\ \mathrm{M_{\odot}}$. This total value is in broad agreement with recent estimates based on red giant stars in \emph{Gaia} DR2. We determine that the mass below $e = 0.7$ is $M_{*,e < 0.7} = 8.6^{+2.4}_{-1.7}\times10^{8}\ \mathrm{M_{\odot}}$, whereas $M_{*,e > 0.7} = 4.6^{+0.7}_{-0.6}\times10^{8}\ \mathrm{M_{\odot}}$. This implies a fraction of mass at $e > 0.7$ of $\sim 34\%$. This fraction changes significantly depending on the cut in $e$, but mainly because lower $e$ limits include more high $e$, high \mgfe{} stars, which are unlikely to be part of the accreted halo (see Figure \ref{fig:mass_MAPs}, and Section \ref{sec:lowmass}). We demonstrate in Appendix \ref{sec:hst} that our mass estimate is robust against systematics introduced by our application of the PARSEC models.
    \item The distribution of mass as a function of \feh{} and \mgfe{} is quite different at high and low $e$, with more mass contributed by low \mgfe{} populations at high $e$. This is likely because the high $e$ population contains stars from the massive accretion event identified as the \emph{Gaia}-Sausage \citep[e.g.][]{2018MNRAS.478..611B}, and \emph{Gaia}-\emph{Enceladus} \citep[e.g.][]{2018arXiv180606038H}. At low $e$, much of the mass is concentrated at higher \mgfe{} (see Figure \ref{fig:mass_MAPs}).
    \item We propose in Section \ref{sec:break} that by considering the stellar halo as the combination of constituent MAPs, we provide a means to reconcile the variety of results which find different density profiles for the halo. MAPs provide insights into the origin of different components of the stellar halo, and demonstrate, for example, that the halo components with shallow power laws are indeed more likely to arise from accreted components.
    \item We estimate and remove the mass in stars which are likely formed \emph{in-situ} but heated to high $e$ \citep[e.g.][]{2018arXiv181208232D,2019arXiv190904679B} to estimate the mass in accreted stars. We estimate the mass of accreted stars at $e > 0.7$ to be $M_{*,\mathrm{accreted}} = 3\pm 1\ \mathrm{(stat.)} \pm 1\ \mathrm{(syst.)}\times10^{8}\ \mathrm{M_\odot}$. If the majority of this population is contributed by the massive accreted dwarf identified by e.g. \citep{2018arXiv180606038H,2018MNRAS.478..611B,2018arXiv180800968M,2019MNRAS.488.1235M}, then this sets an upper limit to its mass which is much lower than the majority of the currently proposed estimates, and implies a total halo mass of the progenitor of $M_{200} \sim 10^{10.2\pm 0.2}\ \mathrm{M_\odot}$ \citep[based on the stellar-halo mass relationship of][]{2017MNRAS.467.2019R}. 
    \item Again by removing the MAPs which we expect to be dominated by \emph{in-situ} stars, we place an estimate of the mass of accreted stars in the halo, independent of kinematics. We find that $M_{*,\mathrm{accreted}} = 9^{+2}_{-1}\times 10^{8}\ \mathrm{M_{\odot}}$, which equates to an accreted fraction $f_{\mathrm{accreted}} = 0.7\pm0.1$, suggesting that the stellar halo is made up mainly of accreted stellar populations. Of this mass, the \emph{Gaia-Enceladus}/Sausage progenitor makes up between 30 and 50$\%$.
\end{itemize}

In this paper we have made a novel mapping of the Milky Way stellar halo, determining not only its total mass inside 70 kpc and between $-3 < \feh{} < -1$ and $0.0 < \mgfe{} < 0.4$, but also studying the distribution of that mass in the space of \feh{}, \mgfe{} and $e$. We have demonstrated that the spatial structure of the stellar halo changes significantly depending on the MAP considered, and as a function of eccentricity. 

Our estimate of the mass of the massive accreted dwarf which forms the majority of the radially biased halo population, and is commonly referred to as \emph{Gaia-Enceladus} or the \emph{Gaia}-Sausage, is not only a key constraint towards understanding the nature of this stellar population -- which is now thought to have played a leading role in the history of the Galaxy -- but is an important point on the mass assembly history of the Milky Way. In the advent of more extensive data in the Milky Way halo \citep[e.g.][]{2019arXiv190707684C}, the methodology we employed here could be used with higher resolution in kinematics and element abundance space, to better map the mass distribution of the halo. Furthermore, in the event that stellar ages become available for large enough samples of halo stars, one could conceivably make time-resolved constraints on the halo mass assembly, which themselves would be a key constraint towards understanding the cosmological context of the Galaxy.

\section*{Acknowledgements}

We thank the anonymous reviewer for a useful report. We also kindly thank Peter Frinchaboy, John Donor, Jon Holtzman, Drew Chojnowski and Gail Zasowski for support in gathering the necessary information for the APOGEE-2 selection function, and Michele Bellazzini, Emanuele Dalessandro and Giacomo Beccari for provision of their HST photometry of NGC2419. We thank Ricardo P. Schiavon, Andrea Miglio, James Lane and Danny Horta-Darrington for useful comments on the original manuscript and warmly thank Vasily Belokurov, Alis Deason, N. Wyn Evans, Giuliano Iorio, Lachlan Lancaster and Jason Sanders for helpful discussions during the preparation of this work. JTM acknowledges support from the ERC Consolidator Grant funding scheme (project ASTEROCHRONOMETRY, G.A. n. 772293), and thanks the Dunlap Institute, CITA and the Department of Astronomy at the University of Toronto for their hospitality during the preparation of the paper. JB received support from the Natural Sciences and Engineering Research Council of Canada (NSERC; funding reference number RGPIN-2015-05235).

This research made use of the cross-match service provided
by CDS, Strasbourg. Analyses and plots presented in this article
used \texttt{iPython}, and packages in the \texttt{SciPy} ecosystem
\citep{Jones:2001aa,4160265,4160251,5725236}.

Funding for the Sloan Digital Sky Survey IV has been provided by the Alfred P. Sloan Foundation, the U.S. Department of Energy Office of Science, and the Participating Institutions. SDSS-IV acknowledges
support and resources from the Center for High-Performance Computing at
the University of Utah. The SDSS web site is www.sdss.org.

SDSS-IV is managed by the Astrophysical Research Consortium for the 
Participating Institutions of the SDSS Collaboration including the 
Brazilian Participation Group, the Carnegie Institution for Science, 
Carnegie Mellon University, the Chilean Participation Group, the French Participation Group, Harvard-Smithsonian Center for Astrophysics, 
Instituto de Astrof\'isica de Canarias, The Johns Hopkins University, Kavli Institute for the Physics and Mathematics of the Universe (IPMU) / 
University of Tokyo, the Korean Participation Group, Lawrence Berkeley National Laboratory, 
Leibniz Institut f\"ur Astrophysik Potsdam (AIP),  
Max-Planck-Institut f\"ur Astronomie (MPIA Heidelberg), 
Max-Planck-Institut f\"ur Astrophysik (MPA Garching), 
Max-Planck-Institut f\"ur Extraterrestrische Physik (MPE), 
National Astronomical Observatories of China, New Mexico State University, 
New York University, University of Notre Dame, 
Observat\'ario Nacional / MCTI, The Ohio State University, 
Pennsylvania State University, Shanghai Astronomical Observatory, 
United Kingdom Participation Group,
Universidad Nacional Aut\'onoma de M\'exico, University of Arizona, 
University of Colorado Boulder, University of Oxford, University of Portsmouth, 
University of Utah, University of Virginia, University of Washington, University of Wisconsin, 
Vanderbilt University, and Yale University.

This work has made use of data from the European Space Agency (ESA) mission
{\it Gaia} (\url{https://www.cosmos.esa.int/gaia}), processed by the {\it Gaia}
Data Processing and Analysis Consortium (DPAC,
\url{https://www.cosmos.esa.int/web/gaia/dpac/consortium}). Funding for the DPAC
has been provided by national institutions, in particular the institutions
participating in the {\it Gaia} Multilateral Agreement.




\bibliographystyle{mnras}
\bibliography{bib} 

\begin{thebibliography}{}
\makeatletter
\relax
\def\mn@urlcharsother{\let\do\@makeother \do\$\do\&\do\#\do\^\do\_\do\%\do\~}
\def\mn@doi{\begingroup\mn@urlcharsother \@ifnextchar [ {\mn@doi@}
  {\mn@doi@[]}}
\def\mn@doi@[#1]#2{\def\@tempa{#1}\ifx\@tempa\@empty \href
  {http://dx.doi.org/#2} {doi:#2}\else \href {http://dx.doi.org/#2} {#1}\fi
  \endgroup}
\def\mn@eprint#1#2{\mn@eprint@#1:#2::\@nil}
\def\mn@eprint@arXiv#1{\href {http://arxiv.org/abs/#1} {{\tt arXiv:#1}}}
\def\mn@eprint@dblp#1{\href {http://dblp.uni-trier.de/rec/bibtex/#1.xml}
  {dblp:#1}}
\def\mn@eprint@#1:#2:#3:#4\@nil{\def\@tempa {#1}\def\@tempb {#2}\def\@tempc
  {#3}\ifx \@tempc \@empty \let \@tempc \@tempb \let \@tempb \@tempa \fi \ifx
  \@tempb \@empty \def\@tempb {arXiv}\fi \@ifundefined
  {mn@eprint@\@tempb}{\@tempb:\@tempc}{\expandafter \expandafter \csname
  mn@eprint@\@tempb\endcsname \expandafter{\@tempc}}}

\bibitem[\protect\citeauthoryear{{Abolfathi} et~al.,}{{Abolfathi}
  et~al.}{2018}]{2018ApJS..235...42A}
{Abolfathi} B.,  et~al., 2018, \mn@doi [\apjs] {10.3847/1538-4365/aa9e8a},
  \href {http://adsabs.harvard.edu/abs/2018ApJS..235...42A} {235, 42}

\bibitem[\protect\citeauthoryear{{Barb{\'a}}, {Minniti}, {Geisler},
  {Alonso-Garc{\'\i}a}, {Hempel}, {Monachesi}, {Arias}  \&
  {G{\'o}mez}}{{Barb{\'a}} et~al.}{2019}]{2019ApJ...870L..24B}
{Barb{\'a}} R.~H.,  {Minniti} D.,  {Geisler} D.,  {Alonso-Garc{\'\i}a} J.,
  {Hempel} M.,  {Monachesi} A.,  {Arias} J.~I.,   {G{\'o}mez} F.~A.,  2019,
  \mn@doi [\apjl] {10.3847/2041-8213/aaf811}, \href
  {https://ui.adsabs.harvard.edu/abs/2019ApJ...870L..24B} {870, L24}

\bibitem[\protect\citeauthoryear{{Baumgardt} \& {Hilker}}{{Baumgardt} \&
  {Hilker}}{2018}]{2018MNRAS.478.1520B}
{Baumgardt} H.,  {Hilker} M.,  2018, \mn@doi [\mnras] {10.1093/mnras/sty1057},
  \href {https://ui.adsabs.harvard.edu/abs/2018MNRAS.478.1520B} {478, 1520}

\bibitem[\protect\citeauthoryear{{Bell} et~al.,}{{Bell}
  et~al.}{2008}]{2008ApJ...680..295B}
{Bell} E.~F.,  et~al., 2008, \mn@doi [\apj] {10.1086/588032}, \href
  {https://ui.adsabs.harvard.edu/abs/2008ApJ...680..295B} {680, 295}

\bibitem[\protect\citeauthoryear{{Bellazzini}, {Dalessandro}, {Sollima}  \&
  {Ibata}}{{Bellazzini} et~al.}{2012}]{2012MNRAS.423..844B}
{Bellazzini} M.,  {Dalessandro} E.,  {Sollima} A.,   {Ibata} R.,  2012, \mn@doi
  [\mnras] {10.1111/j.1365-2966.2012.20922.x}, \href
  {https://ui.adsabs.harvard.edu/abs/2012MNRAS.423..844B} {423, 844}

\bibitem[\protect\citeauthoryear{{Belokurov}, {Erkal}, {Evans}, {Koposov}  \&
  {Deason}}{{Belokurov} et~al.}{2018}]{2018MNRAS.478..611B}
{Belokurov} V.,  {Erkal} D.,  {Evans} N.~W.,  {Koposov} S.~E.,   {Deason}
  A.~J.,  2018, \mn@doi [\mnras] {10.1093/mnras/sty982}, \href
  {http://adsabs.harvard.edu/abs/2018MNRAS.478..611B} {478, 611}

\bibitem[\protect\citeauthoryear{{Belokurov}, {Sanders}, {Fattahi}, {Smith},
  {Deason}, {Evans}  \& {Grand }}{{Belokurov}
  et~al.}{2019}]{2019arXiv190904679B}
{Belokurov} V.,  {Sanders} J.~L.,  {Fattahi} A.,  {Smith} M.~C.,  {Deason}
  A.~J.,  {Evans} N.~W.,   {Grand } R. J.~J.,  2019, arXiv e-prints, \href
  {https://ui.adsabs.harvard.edu/abs/2019arXiv190904679B} {p. arXiv:1909.04679}

\bibitem[\protect\citeauthoryear{{Bennett} \& {Bovy}}{{Bennett} \&
  {Bovy}}{2018}]{2018arXiv180903507B}
{Bennett} M.,  {Bovy} J.,  2018, preprint, \href
  {http://adsabs.harvard.edu/abs/2018arXiv180903507B} {} (\mn@eprint {arXiv}
  {1809.03507})

\bibitem[\protect\citeauthoryear{{Bland-Hawthorn} \&
  {Gerhard}}{{Bland-Hawthorn} \& {Gerhard}}{2016}]{2016ARA&A..54..529B}
{Bland-Hawthorn} J.,  {Gerhard} O.,  2016, \mn@doi [\araa]
  {10.1146/annurev-astro-081915-023441}, \href
  {http://ukads.nottingham.ac.uk/abs/2016ARA%26A..54..529B} {54, 529}

\bibitem[\protect\citeauthoryear{{Bland-Hawthorn} et~al.,}{{Bland-Hawthorn}
  et~al.}{2018}]{2018arXiv180902658B}
{Bland-Hawthorn} J.,  et~al., 2018, preprint, \href
  {http://adsabs.harvard.edu/abs/2018arXiv180902658B} {} (\mn@eprint {arXiv}
  {1809.02658})

\bibitem[\protect\citeauthoryear{{Blanton} et~al.,}{{Blanton}
  et~al.}{2017}]{2017AJ....154...28B}
{Blanton} M.~R.,  et~al., 2017, \mn@doi [\aj] {10.3847/1538-3881/aa7567}, \href
  {http://adsabs.harvard.edu/abs/2017AJ....154...28B} {154, 28}

\bibitem[\protect\citeauthoryear{{Bovy}}{{Bovy}}{2015}]{2015ApJS..216...29B}
{Bovy} J.,  2015, \mn@doi [\apjs] {10.1088/0067-0049/216/2/29}, \href
  {http://adsabs.harvard.edu/abs/2015ApJS..216...29B} {216, 29}

\bibitem[\protect\citeauthoryear{{Bovy} et~al.,}{{Bovy}
  et~al.}{2014}]{2014ApJ...790..127B}
{Bovy} J.,  et~al., 2014, \mn@doi [\apj] {10.1088/0004-637X/790/2/127}, \href
  {http://adsabs.harvard.edu/abs/2014ApJ...790..127B} {790, 127}

\bibitem[\protect\citeauthoryear{{Bovy}, {Rix}, {Green}, {Schlafly}  \&
  {Finkbeiner}}{{Bovy} et~al.}{2016a}]{2016ApJ...818..130B}
{Bovy} J.,  {Rix} H.-W.,  {Green} G.~M.,  {Schlafly} E.~F.,   {Finkbeiner}
  D.~P.,  2016a, \mn@doi [\apj] {10.3847/0004-637X/818/2/130}, \href
  {http://adsabs.harvard.edu/abs/2016ApJ...818..130B} {818, 130}

\bibitem[\protect\citeauthoryear{{Bovy}, {Rix}, {Schlafly}, {Nidever},
  {Holtzman}, {Shetrone}  \& {Beers}}{{Bovy}
  et~al.}{2016b}]{2016ApJ...823...30B}
{Bovy} J.,  {Rix} H.-W.,  {Schlafly} E.~F.,  {Nidever} D.~L.,  {Holtzman}
  J.~A.,  {Shetrone} M.,   {Beers} T.~C.,  2016b, \mn@doi [\apj]
  {10.3847/0004-637X/823/1/30}, \href
  {http://adsabs.harvard.edu/abs/2016ApJ...823...30B} {823, 30}

\bibitem[\protect\citeauthoryear{{Bovy}, {Bahmanyar}, {Fritz}  \&
  {Kallivayalil}}{{Bovy} et~al.}{2016c}]{2016ApJ...833...31B}
{Bovy} J.,  {Bahmanyar} A.,  {Fritz} T.~K.,   {Kallivayalil} N.,  2016c,
  \mn@doi [\apj] {10.3847/1538-4357/833/1/31}, \href
  {https://ui.adsabs.harvard.edu/abs/2016ApJ...833...31B} {833, 31}

\bibitem[\protect\citeauthoryear{{Bovy}, {Erkal}  \& {Sanders}}{{Bovy}
  et~al.}{2017}]{2017MNRAS.466..628B}
{Bovy} J.,  {Erkal} D.,   {Sanders} J.~L.,  2017, \mn@doi [\mnras]
  {10.1093/mnras/stw3067}, \href
  {https://ui.adsabs.harvard.edu/abs/2017MNRAS.466..628B} {466, 628}

\bibitem[\protect\citeauthoryear{{Bressan}, {Marigo}, {Girardi}, {Salasnich},
  {Dal Cero}, {Rubele}  \& {Nanni}}{{Bressan}
  et~al.}{2012}]{2012MNRAS.427..127B}
{Bressan} A.,  {Marigo} P.,  {Girardi} L.,  {Salasnich} B.,  {Dal Cero} C.,
  {Rubele} S.,   {Nanni} A.,  2012, \mn@doi [\mnras]
  {10.1111/j.1365-2966.2012.21948.x}, \href
  {http://adsabs.harvard.edu/abs/2012MNRAS.427..127B} {427, 127}

\bibitem[\protect\citeauthoryear{{Bullock} \& {Johnston}}{{Bullock} \&
  {Johnston}}{2005}]{2005ApJ...635..931B}
{Bullock} J.~S.,  {Johnston} K.~V.,  2005, \mn@doi [\apj] {10.1086/497422},
  \href {https://ui.adsabs.harvard.edu/abs/2005ApJ...635..931B} {635, 931}

\bibitem[\protect\citeauthoryear{{Carollo} et~al.,}{{Carollo}
  et~al.}{2007}]{2007Natur.450.1020C}
{Carollo} D.,  et~al., 2007, \mn@doi [\nat] {10.1038/nature06460}, \href
  {http://adsabs.harvard.edu/abs/2007Natur.450.1020C} {450, 1020}

\bibitem[\protect\citeauthoryear{{Conroy} et~al.,}{{Conroy}
  et~al.}{2019}]{2019arXiv190707684C}
{Conroy} C.,  et~al., 2019, arXiv e-prints, \href
  {https://ui.adsabs.harvard.edu/abs/2019arXiv190707684C} {p. arXiv:1907.07684}

\bibitem[\protect\citeauthoryear{{Crain} et~al.,}{{Crain}
  et~al.}{2015}]{2015MNRAS.450.1937C}
{Crain} R.~A.,  et~al., 2015, \mn@doi [\mnras] {10.1093/mnras/stv725}, \href
  {http://adsabs.harvard.edu/abs/2015MNRAS.450.1937C} {450, 1937}

\bibitem[\protect\citeauthoryear{{Das}, {Hawkins}  \& {Jofre}}{{Das}
  et~al.}{2019}]{2019arXiv190309320D}
{Das} P.,  {Hawkins} K.,   {Jofre} P.,  2019, arXiv e-prints, \href
  {https://ui.adsabs.harvard.edu/abs/2019arXiv190309320D} {p. arXiv:1903.09320}

\bibitem[\protect\citeauthoryear{{Deason}, {Belokurov}  \& {Evans}}{{Deason}
  et~al.}{2011}]{2011MNRAS.416.2903D}
{Deason} A.~J.,  {Belokurov} V.,   {Evans} N.~W.,  2011, \mn@doi [\mnras]
  {10.1111/j.1365-2966.2011.19237.x}, \href
  {https://ui.adsabs.harvard.edu/abs/2011MNRAS.416.2903D} {416, 2903}

\bibitem[\protect\citeauthoryear{{Deason}, {Belokurov}, {Evans}  \&
  {Johnston}}{{Deason} et~al.}{2013}]{2013ApJ...763..113D}
{Deason} A.~J.,  {Belokurov} V.,  {Evans} N.~W.,   {Johnston} K.~V.,  2013,
  \mn@doi [\apj] {10.1088/0004-637X/763/2/113}, \href
  {http://adsabs.harvard.edu/abs/2013ApJ...763..113D} {763, 113}

\bibitem[\protect\citeauthoryear{{Deason}, {Belokurov}  \& {Sanders}}{{Deason}
  et~al.}{2019}]{2019arXiv190802763D}
{Deason} A.~J.,  {Belokurov} V.,   {Sanders} J.~L.,  2019, arXiv e-prints,
  \href {https://ui.adsabs.harvard.edu/abs/2019arXiv190802763D} {p.
  arXiv:1908.02763}

\bibitem[\protect\citeauthoryear{{Di Matteo}, {Haywood}, {Lehnert}, {Katz},
  {Khoperskov}, {Snaith}, {G{\'o}mez}  \& {Robichon}}{{Di Matteo}
  et~al.}{2018}]{2018arXiv181208232D}
{Di Matteo} P.,  {Haywood} M.,  {Lehnert} M.~D.,  {Katz} D.,  {Khoperskov} S.,
  {Snaith} O.~N.,  {G{\'o}mez} A.,   {Robichon} N.,  2018, arXiv e-prints,
  \href {https://ui.adsabs.harvard.edu/abs/2018arXiv181208232D} {p.
  arXiv:1812.08232}

\bibitem[\protect\citeauthoryear{{Digby}, {Hambly}, {Cooke}, {Reid}  \&
  {Cannon}}{{Digby} et~al.}{2003}]{2003MNRAS.344..583D}
{Digby} A.~P.,  {Hambly} N.~C.,  {Cooke} J.~A.,  {Reid} I.~N.,   {Cannon}
  R.~D.,  2003, \mn@doi [\mnras] {10.1046/j.1365-8711.2003.06842.x}, \href
  {https://ui.adsabs.harvard.edu/abs/2003MNRAS.344..583D} {344, 583}

\bibitem[\protect\citeauthoryear{{Drimmel}, {Cabrera-Lavers}  \&
  {L{\'o}pez-Corredoira}}{{Drimmel} et~al.}{2003}]{2003A&A...409..205D}
{Drimmel} R.,  {Cabrera-Lavers} A.,   {L{\'o}pez-Corredoira} M.,  2003, \mn@doi
  [\aap] {10.1051/0004-6361:20031070}, \href
  {https://ui.adsabs.harvard.edu/abs/2003A%26A...409..205D} {409, 205}

\bibitem[\protect\citeauthoryear{{Einasto}}{{Einasto}}{1965}]{1965TrAlm...5...87E}
{Einasto} J.,  1965, Trudy Astrofizicheskogo Instituta Alma-Ata, \href
  {https://ui.adsabs.harvard.edu/abs/1965TrAlm...5...87E} {5, 87}

\bibitem[\protect\citeauthoryear{{Eisenstein} et~al.,}{{Eisenstein}
  et~al.}{2011}]{2011AJ....142...72E}
{Eisenstein} D.~J.,  et~al., 2011, \mn@doi [\aj] {10.1088/0004-6256/142/3/72},
  \href {http://adsabs.harvard.edu/abs/2011AJ....142...72E} {142, 72}

\bibitem[\protect\citeauthoryear{{Fattahi} et~al.,}{{Fattahi}
  et~al.}{2019}]{2019MNRAS.484.4471F}
{Fattahi} A.,  et~al., 2019, \mn@doi [\mnras] {10.1093/mnras/stz159}, \href
  {https://ui.adsabs.harvard.edu/abs/2019MNRAS.484.4471F} {484, 4471}

\bibitem[\protect\citeauthoryear{{Fern{\'a}ndez-Alvar}
  et~al.,}{{Fern{\'a}ndez-Alvar} et~al.}{2018}]{2018ApJ...852...50F}
{Fern{\'a}ndez-Alvar} E.,  et~al., 2018, \mn@doi [\apj]
  {10.3847/1538-4357/aa9ced}, \href
  {http://adsabs.harvard.edu/abs/2018ApJ...852...50F} {852, 50}

\bibitem[\protect\citeauthoryear{{Font}, {McCarthy}, {Crain}, {Theuns},
  {Schaye}, {Wiersma}  \& {Dalla Vecchia}}{{Font}
  et~al.}{2011}]{2011MNRAS.416.2802F}
{Font} A.~S.,  {McCarthy} I.~G.,  {Crain} R.~A.,  {Theuns} T.,  {Schaye} J.,
  {Wiersma} R.~P.~C.,   {Dalla Vecchia} C.,  2011, \mn@doi [\mnras]
  {10.1111/j.1365-2966.2011.19227.x}, \href
  {https://ui.adsabs.harvard.edu/abs/2011MNRAS.416.2802F} {416, 2802}

\bibitem[\protect\citeauthoryear{{Forbes} \& {Bridges}}{{Forbes} \&
  {Bridges}}{2010}]{2010MNRAS.404.1203F}
{Forbes} D.~A.,  {Bridges} T.,  2010, \mn@doi [\mnras]
  {10.1111/j.1365-2966.2010.16373.x}, \href
  {https://ui.adsabs.harvard.edu/abs/2010MNRAS.404.1203F} {404, 1203}

\bibitem[\protect\citeauthoryear{{Foreman-Mackey}, {Hogg}, {Lang}  \&
  {Goodman}}{{Foreman-Mackey} et~al.}{2013}]{2013PASP..125..306F}
{Foreman-Mackey} D.,  {Hogg} D.~W.,  {Lang} D.,   {Goodman} J.,  2013, \mn@doi
  [\pasp] {10.1086/670067}, \href
  {http://adsabs.harvard.edu/abs/2013PASP..125..306F} {125, 306}

\bibitem[\protect\citeauthoryear{{Fuchs} \& {Jahrei{\ss}}}{{Fuchs} \&
  {Jahrei{\ss}}}{1998}]{1998A&A...329...81F}
{Fuchs} B.,  {Jahrei{\ss}} H.,  1998, \aap, \href
  {https://ui.adsabs.harvard.edu/abs/1998A&A...329...81F} {329, 81}

\bibitem[\protect\citeauthoryear{{Gaia Collaboration}, {Brown}, {Vallenari},
  {Prusti}, {de Bruijne}, {Babusiaux}  \& {Bailer-Jones}}{{Gaia Collaboration}
  et~al.}{2018}]{2018arXiv180409365G}
{Gaia Collaboration} {Brown} A.~G.~A.,  {Vallenari} A.,  {Prusti} T.,  {de
  Bruijne} J.~H.~J.,  {Babusiaux} C.,   {Bailer-Jones} C.~A.~L.,  2018,
  preprint, \href {http://adsabs.harvard.edu/abs/2018arXiv180409365G} {}
  (\mn@eprint {arXiv} {1804.09365})

\bibitem[\protect\citeauthoryear{{Garc{\'{\i}}a P{\'e}rez}
  et~al.,}{{Garc{\'{\i}}a P{\'e}rez} et~al.}{2016}]{2016AJ....151..144G}
{Garc{\'{\i}}a P{\'e}rez} A.~E.,  et~al., 2016, \mn@doi [\aj]
  {10.3847/0004-6256/151/6/144}, \href
  {http://adsabs.harvard.edu/abs/2016AJ....151..144G} {151, 144}

\bibitem[\protect\citeauthoryear{{Geisler}, {Wallerstein}, {Smith}  \&
  {Casetti-Dinescu}}{{Geisler} et~al.}{2007}]{2007PASP..119..939G}
{Geisler} D.,  {Wallerstein} G.,  {Smith} V.~V.,   {Casetti-Dinescu} D.~I.,
  2007, \mn@doi [\pasp] {10.1086/521990}, \href
  {https://ui.adsabs.harvard.edu/abs/2007PASP..119..939G} {119, 939}

\bibitem[\protect\citeauthoryear{Goodman \& Weare}{Goodman \&
  Weare}{2010}]{goodmanweare2010}
Goodman J.,  Weare J.,  2010, Comm. App. Math. and Comp. Sci., 65

\bibitem[\protect\citeauthoryear{{Gould}, {Flynn}  \& {Bahcall}}{{Gould}
  et~al.}{1998}]{1998ApJ...503..798G}
{Gould} A.,  {Flynn} C.,   {Bahcall} J.~N.,  1998, \mn@doi [\apj]
  {10.1086/306023}, \href
  {https://ui.adsabs.harvard.edu/abs/1998ApJ...503..798G} {503, 798}

\bibitem[\protect\citeauthoryear{{Grand} et~al.,}{{Grand}
  et~al.}{2017}]{2017MNRAS.467..179G}
{Grand} R.~J.~J.,  et~al., 2017, \mn@doi [\mnras] {10.1093/mnras/stx071}, \href
  {http://adsabs.harvard.edu/abs/2017MNRAS.467..179G} {467, 179}

\bibitem[\protect\citeauthoryear{{Gratton}, {D'Orazi}, {Bragaglia}, {Carretta}
  \& {Lucatello}}{{Gratton} et~al.}{2010}]{2010A&A...522A..77G}
{Gratton} R.~G.,  {D'Orazi} V.,  {Bragaglia} A.,  {Carretta} E.,   {Lucatello}
  S.,  2010, \mn@doi [\aap] {10.1051/0004-6361/201015405}, \href
  {https://ui.adsabs.harvard.edu/abs/2010A&A...522A..77G} {522, A77}

\bibitem[\protect\citeauthoryear{{Gravity Collaboration} et~al.,}{{Gravity
  Collaboration} et~al.}{2018}]{2018A&A...615L..15G}
{Gravity Collaboration} et~al., 2018, \mn@doi [\aap]
  {10.1051/0004-6361/201833718}, \href
  {https://ui.adsabs.harvard.edu/\#abs/2018A&A...615L..15G} {615, L15}

\bibitem[\protect\citeauthoryear{{Green} et~al.,}{{Green}
  et~al.}{2015}]{2015ApJ...810...25G}
{Green} G.~M.,  et~al., 2015, \mn@doi [\apj] {10.1088/0004-637X/810/1/25},
  \href {http://adsabs.harvard.edu/abs/2015ApJ...810...25G} {810, 25}

\bibitem[\protect\citeauthoryear{{Gunn} et~al.,}{{Gunn}
  et~al.}{2006}]{2006AJ....131.2332G}
{Gunn} J.~E.,  et~al., 2006, \mn@doi [\aj] {10.1086/500975}, \href
  {http://adsabs.harvard.edu/abs/2006AJ....131.2332G} {131, 2332}

\bibitem[\protect\citeauthoryear{{Harris}}{{Harris}}{1996}]{1996AJ....112.1487H}
{Harris} W.~E.,  1996, \mn@doi [\aj] {10.1086/118116}, \href
  {http://adsabs.harvard.edu/abs/1996AJ....112.1487H} {112, 1487}

\bibitem[\protect\citeauthoryear{{Hayes} et~al.,}{{Hayes}
  et~al.}{2018}]{2018ApJ...852...49H}
{Hayes} C.~R.,  et~al., 2018, \mn@doi [\apj] {10.3847/1538-4357/aa9cec}, \href
  {http://adsabs.harvard.edu/abs/2018ApJ...852...49H} {852, 49}

\bibitem[\protect\citeauthoryear{{Helmi}, {Babusiaux}, {Koppelman}, {Massari},
  {Veljanoski}  \& {Brown}}{{Helmi} et~al.}{2018}]{2018arXiv180606038H}
{Helmi} A.,  {Babusiaux} C.,  {Koppelman} H.~H.,  {Massari} D.,  {Veljanoski}
  J.,   {Brown} A.~G.~A.,  2018, preprint, \href
  {http://adsabs.harvard.edu/abs/2018arXiv180606038H} {} (\mn@eprint {arXiv}
  {1806.06038})

\bibitem[\protect\citeauthoryear{{Holtzman} et~al.,}{{Holtzman}
  et~al.}{2018}]{holtzdr14}
{Holtzman} J.,  et~al., 2018, AJ

\bibitem[\protect\citeauthoryear{Hunter}{Hunter}{2007}]{4160265}
Hunter J.~D.,  2007, \mn@doi [Computing in Science Engineering]
  {10.1109/MCSE.2007.55}, 9, 90

\bibitem[\protect\citeauthoryear{{Iorio} \& {Belokurov}}{{Iorio} \&
  {Belokurov}}{2019}]{2019MNRAS.482.3868I}
{Iorio} G.,  {Belokurov} V.,  2019, \mn@doi [\mnras] {10.1093/mnras/sty2806},
  \href {https://ui.adsabs.harvard.edu/abs/2019MNRAS.482.3868I} {482, 3868}

\bibitem[\protect\citeauthoryear{{Iorio}, {Belokurov}, {Erkal}, {Koposov},
  {Nipoti}  \& {Fraternali}}{{Iorio} et~al.}{2018}]{2018MNRAS.474.2142I}
{Iorio} G.,  {Belokurov} V.,  {Erkal} D.,  {Koposov} S.~E.,  {Nipoti} C.,
  {Fraternali} F.,  2018, \mn@doi [\mnras] {10.1093/mnras/stx2819}, \href
  {https://ui.adsabs.harvard.edu/abs/2018MNRAS.474.2142I} {474, 2142}

\bibitem[\protect\citeauthoryear{Jones, Oliphant, Peterson  et~al.}{Jones
  et~al.}{2001}]{Jones:2001aa}
Jones E.,  Oliphant T.,  Peterson P.,   et~al., 2001, {SciPy}: Open source
  scientific tools for {Python}, \url {http://www.scipy.org/}

\bibitem[\protect\citeauthoryear{{J\"onsson} et~al.,}{{J\"onsson}
  et~al.}{2018}]{jonssondr14}
{J\"onsson} H.,  et~al., 2018, AJ

\bibitem[\protect\citeauthoryear{{Juri{\'c}} et~al.,}{{Juri{\'c}}
  et~al.}{2008}]{2008ApJ...673..864J}
{Juri{\'c}} M.,  et~al., 2008, \mn@doi [\apj] {10.1086/523619}, \href
  {http://adsabs.harvard.edu/abs/2008ApJ...673..864J} {673, 864}

\bibitem[\protect\citeauthoryear{{Kroupa}}{{Kroupa}}{2001}]{2001MNRAS.322..231K}
{Kroupa} P.,  2001, \mn@doi [\mnras] {10.1046/j.1365-8711.2001.04022.x}, \href
  {http://adsabs.harvard.edu/abs/2001MNRAS.322..231K} {322, 231}

\bibitem[\protect\citeauthoryear{{Kruijssen}, {Pfeffer}, {Reina-Campos},
  {Crain}  \& {Bastian}}{{Kruijssen} et~al.}{2018}]{2018MNRAS.tmp.1537K}
{Kruijssen} J.~M.~D.,  {Pfeffer} J.~L.,  {Reina-Campos} M.,  {Crain} R.~A.,
  {Bastian} N.,  2018, \mn@doi [\mnras] {10.1093/mnras/sty1609}, \href
  {http://adsabs.harvard.edu/abs/2018MNRAS.tmp.1537K} {}

\bibitem[\protect\citeauthoryear{{Lancaster}, {Koposov}, {Belokurov}, {Evans}
  \& {Deason}}{{Lancaster} et~al.}{2019}]{2019MNRAS.486..378L}
{Lancaster} L.,  {Koposov} S.~E.,  {Belokurov} V.,  {Evans} N.~W.,   {Deason}
  A.~J.,  2019, \mn@doi [\mnras] {10.1093/mnras/stz853}, \href
  {https://ui.adsabs.harvard.edu/abs/2019MNRAS.486..378L} {486, 378}

\bibitem[\protect\citeauthoryear{{Leung} \& {Bovy}}{{Leung} \&
  {Bovy}}{2019a}]{2019arXiv190208634L}
{Leung} H.~W.,  {Bovy} J.,  2019a, arXiv e-prints, \href
  {https://ui.adsabs.harvard.edu/\#abs/2019arXiv190208634L} {p.
  arXiv:1902.08634}

\bibitem[\protect\citeauthoryear{{Leung} \& {Bovy}}{{Leung} \&
  {Bovy}}{2019b}]{2018arXiv180804428L}
{Leung} H.~W.,  {Bovy} J.,  2019b, \mn@doi [\mnras] {10.1093/mnras/sty3217},
  \href {https://ui.adsabs.harvard.edu/\#abs/2019MNRAS.483.3255L} {483, 3255}

\bibitem[\protect\citeauthoryear{{Mackereth} \& {Bovy}}{{Mackereth} \&
  {Bovy}}{2018}]{2018arXiv180202592M}
{Mackereth} J.~T.,  {Bovy} J.,  2018, preprint, \href
  {http://adsabs.harvard.edu/abs/2018arXiv180202592M} {} (\mn@eprint {arXiv}
  {1802.02592})

\bibitem[\protect\citeauthoryear{{Mackereth} et~al.,}{{Mackereth}
  et~al.}{2017}]{2017arXiv170600018M}
{Mackereth} J.~T.,  et~al., 2017, \mn@doi [\mnras] {10.1093/mnras/stx1774},
  \href {http://adsabs.harvard.edu/abs/2017MNRAS.471.3057M} {471, 3057}

\bibitem[\protect\citeauthoryear{{Mackereth} et~al.,}{{Mackereth}
  et~al.}{2018a}]{2018arXiv180800968M}
{Mackereth} J.~T.,  et~al., 2018a, preprint, \href
  {http://adsabs.harvard.edu/abs/2018arXiv180800968M} {} (\mn@eprint {arXiv}
  {1808.00968})

\bibitem[\protect\citeauthoryear{{Mackereth}, {Crain}, {Schiavon}, {Schaye},
  {Theuns}  \& {Schaller}}{{Mackereth} et~al.}{2018b}]{2018MNRAS.477.5072M}
{Mackereth} J.~T.,  {Crain} R.~A.,  {Schiavon} R.~P.,  {Schaye} J.,  {Theuns}
  T.,   {Schaller} M.,  2018b, \mn@doi [\mnras] {10.1093/mnras/sty972}, \href
  {http://adsabs.harvard.edu/abs/2018MNRAS.477.5072M} {477, 5072}

\bibitem[\protect\citeauthoryear{{Majewski} et~al.,}{{Majewski}
  et~al.}{2017}]{2015arXiv150905420M}
{Majewski} S.~R.,  et~al., 2017, \mn@doi [\aj] {10.3847/1538-3881/aa784d},
  \href {http://adsabs.harvard.edu/abs/2017AJ....154...94M} {154, 94}

\bibitem[\protect\citeauthoryear{{Marigo} et~al.,}{{Marigo}
  et~al.}{2017}]{2017ApJ...835...77M}
{Marigo} P.,  et~al., 2017, \mn@doi [\apj] {10.3847/1538-4357/835/1/77}, \href
  {https://ui.adsabs.harvard.edu/abs/2017ApJ...835...77M} {835, 77}

\bibitem[\protect\citeauthoryear{{Marshall}, {Robin}, {Reyl{\'e}}, {Schultheis}
   \& {Picaud}}{{Marshall} et~al.}{2006}]{2006A&A...453..635M}
{Marshall} D.~J.,  {Robin} A.~C.,  {Reyl{\'e}} C.,  {Schultheis} M.,   {Picaud}
  S.,  2006, \mn@doi [\aap] {10.1051/0004-6361:20053842}, \href
  {http://adsabs.harvard.edu/abs/2006A%26A...453..635M} {453, 635}

\bibitem[\protect\citeauthoryear{{Mart{\'\i}nez-Delgado}, {Pe{\~n}arrubia},
  {Gabany}, {Trujillo}, {Majewski}  \& {Pohlen}}{{Mart{\'\i}nez-Delgado}
  et~al.}{2008}]{2008ApJ...689..184M}
{Mart{\'\i}nez-Delgado} D.,  {Pe{\~n}arrubia} J.,  {Gabany} R.~J.,  {Trujillo}
  I.,  {Majewski} S.~R.,   {Pohlen} M.,  2008, \mn@doi [\apj] {10.1086/592555},
  \href {https://ui.adsabs.harvard.edu/abs/2008ApJ...689..184M} {689, 184}

\bibitem[\protect\citeauthoryear{{Mart{\'\i}nez-Delgado}, {D'Onghia}, {Chonis},
  {Beaton}, {Teuwen}, {GaBany}, {Grebel}  \& {Morales}}{{Mart{\'\i}nez-Delgado}
  et~al.}{2015}]{2015AJ....150..116M}
{Mart{\'\i}nez-Delgado} D.,  {D'Onghia} E.,  {Chonis} T.~S.,  {Beaton} R.~L.,
  {Teuwen} K.,  {GaBany} R.~J.,  {Grebel} E.~K.,   {Morales} G.,  2015, \mn@doi
  [\aj] {10.1088/0004-6256/150/4/116}, \href
  {https://ui.adsabs.harvard.edu/abs/2015AJ....150..116M} {150, 116}

\bibitem[\protect\citeauthoryear{{McCarthy}, {Font}, {Crain}, {Deason},
  {Schaye}  \& {Theuns}}{{McCarthy} et~al.}{2012}]{2012MNRAS.420.2245M}
{McCarthy} I.~G.,  {Font} A.~S.,  {Crain} R.~A.,  {Deason} A.~J.,  {Schaye} J.,
    {Theuns} T.,  2012, \mn@doi [\mnras] {10.1111/j.1365-2966.2011.20189.x},
  \href {https://ui.adsabs.harvard.edu/abs/2012MNRAS.420.2245M} {420, 2245}

\bibitem[\protect\citeauthoryear{{Morrison}}{{Morrison}}{1993}]{1993AJ....106..578M}
{Morrison} H.~L.,  1993, \mn@doi [\aj] {10.1086/116662}, \href
  {https://ui.adsabs.harvard.edu/abs/1993AJ....106..578M} {106, 578}

\bibitem[\protect\citeauthoryear{{Morrison}, {Mateo}, {Olszewski}, {Harding},
  {Dohm-Palmer}, {Freeman}, {Norris}  \& {Morita}}{{Morrison}
  et~al.}{2000}]{2000AJ....119.2254M}
{Morrison} H.~L.,  {Mateo} M.,  {Olszewski} E.~W.,  {Harding} P.,
  {Dohm-Palmer} R.~C.,  {Freeman} K.~C.,  {Norris} J.~E.,   {Morita} M.,  2000,
  \mn@doi [\aj] {10.1086/301357}, \href
  {https://ui.adsabs.harvard.edu/abs/2000AJ....119.2254M} {119, 2254}

\bibitem[\protect\citeauthoryear{{Myeong}, {Vasiliev}, {Iorio}, {Evans}  \&
  {Belokurov}}{{Myeong} et~al.}{2019}]{2019MNRAS.488.1235M}
{Myeong} G.~C.,  {Vasiliev} E.,  {Iorio} G.,  {Evans} N.~W.,   {Belokurov} V.,
  2019, \mn@doi [\mnras] {10.1093/mnras/stz1770}, \href
  {https://ui.adsabs.harvard.edu/abs/2019MNRAS.488.1235M} {488, 1235}

\bibitem[\protect\citeauthoryear{{Nidever} et~al.,}{{Nidever}
  et~al.}{2015}]{2015AJ....150..173N}
{Nidever} D.~L.,  et~al., 2015, \mn@doi [\aj] {10.1088/0004-6256/150/6/173},
  \href {http://adsabs.harvard.edu/abs/2015AJ....150..173N} {150, 173}

\bibitem[\protect\citeauthoryear{{Nissen} \& {Schuster}}{{Nissen} \&
  {Schuster}}{2010}]{2010A&A...511L..10N}
{Nissen} P.~E.,  {Schuster} W.~J.,  2010, \mn@doi [\aap]
  {10.1051/0004-6361/200913877}, \href
  {http://adsabs.harvard.edu/abs/2010A%26A...511L..10N} {511, L10}

\bibitem[\protect\citeauthoryear{Perez \& Granger}{Perez \&
  Granger}{2007}]{4160251}
Perez F.,  Granger B.~E.,  2007, \mn@doi [Computing in Science Engineering]
  {10.1109/MCSE.2007.53}, 9, 21

\bibitem[\protect\citeauthoryear{{Read}, {Iorio}, {Agertz}  \&
  {Fraternali}}{{Read} et~al.}{2017}]{2017MNRAS.467.2019R}
{Read} J.~I.,  {Iorio} G.,  {Agertz} O.,   {Fraternali} F.,  2017, \mn@doi
  [\mnras] {10.1093/mnras/stx147}, \href
  {https://ui.adsabs.harvard.edu/abs/2017MNRAS.467.2019R} {467, 2019}

\bibitem[\protect\citeauthoryear{{Schaye} et~al.,}{{Schaye}
  et~al.}{2015}]{2015MNRAS.446..521S}
{Schaye} J.,  et~al., 2015, \mn@doi [\mnras] {10.1093/mnras/stu2058}, \href
  {http://adsabs.harvard.edu/abs/2015MNRAS.446..521S} {446, 521}

\bibitem[\protect\citeauthoryear{{Sch{\"o}nrich}, {Binney}  \&
  {Dehnen}}{{Sch{\"o}nrich} et~al.}{2010}]{2010MNRAS.403.1829S}
{Sch{\"o}nrich} R.,  {Binney} J.,   {Dehnen} W.,  2010, \mn@doi [\mnras]
  {10.1111/j.1365-2966.2010.16253.x}, \href
  {http://adsabs.harvard.edu/abs/2010MNRAS.403.1829S} {403, 1829}

\bibitem[\protect\citeauthoryear{{Sch{\"o}nrich}, {Asplund}  \&
  {Casagrande}}{{Sch{\"o}nrich} et~al.}{2011}]{2011MNRAS.415.3807S}
{Sch{\"o}nrich} R.,  {Asplund} M.,   {Casagrande} L.,  2011, \mn@doi [\mnras]
  {10.1111/j.1365-2966.2011.19003.x}, \href
  {https://ui.adsabs.harvard.edu/abs/2011MNRAS.415.3807S} {415, 3807}

\bibitem[\protect\citeauthoryear{{Sch{\"o}nrich}, {Asplund}  \&
  {Casagrande}}{{Sch{\"o}nrich} et~al.}{2014}]{2014ApJ...786....7S}
{Sch{\"o}nrich} R.,  {Asplund} M.,   {Casagrande} L.,  2014, \mn@doi [\apj]
  {10.1088/0004-637X/786/1/7}, \href
  {https://ui.adsabs.harvard.edu/abs/2014ApJ...786....7S} {786, 7}

\bibitem[\protect\citeauthoryear{{Sesar}, {Juri{\'c}}  \& {Ivezi{\'c}}}{{Sesar}
  et~al.}{2011}]{2011ApJ...731....4S}
{Sesar} B.,  {Juri{\'c}} M.,   {Ivezi{\'c}} {\v{Z}}.,  2011, \mn@doi [\apj]
  {10.1088/0004-637X/731/1/4}, \href
  {https://ui.adsabs.harvard.edu/abs/2011ApJ...731....4S} {731, 4}

\bibitem[\protect\citeauthoryear{{Sesar} et~al.,}{{Sesar}
  et~al.}{2013}]{2013AJ....146...21S}
{Sesar} B.,  et~al., 2013, \mn@doi [\aj] {10.1088/0004-6256/146/2/21}, \href
  {https://ui.adsabs.harvard.edu/abs/2013AJ....146...21S} {146, 21}

\bibitem[\protect\citeauthoryear{{Shetrone} et~al.,}{{Shetrone}
  et~al.}{2015}]{2015ApJS..221...24S}
{Shetrone} M.,  et~al., 2015, \mn@doi [\apjs] {10.1088/0067-0049/221/2/24},
  \href {http://adsabs.harvard.edu/abs/2015ApJS..221...24S} {221, 24}

\bibitem[\protect\citeauthoryear{{Skrutskie} et~al.,}{{Skrutskie}
  et~al.}{2006}]{2006AJ....131.1163S}
{Skrutskie} M.~F.,  et~al., 2006, \mn@doi [\aj] {10.1086/498708}, \href
  {http://adsabs.harvard.edu/abs/2006AJ....131.1163S} {131, 1163}

\bibitem[\protect\citeauthoryear{{Spitoni}, {Vincenzo}, {Matteucci}  \&
  {Romano}}{{Spitoni} et~al.}{2016}]{2016MNRAS.458.2541S}
{Spitoni} E.,  {Vincenzo} F.,  {Matteucci} F.,   {Romano} D.,  2016, \mn@doi
  [\mnras] {10.1093/mnras/stw519}, \href
  {https://ui.adsabs.harvard.edu/abs/2016MNRAS.458.2541S} {458, 2541}

\bibitem[\protect\citeauthoryear{{Vincenzo}, {Spitoni}, {Calura}, {Matteucci},
  {Silva Aguirre}, {Miglio}  \& {Cescutti}}{{Vincenzo}
  et~al.}{2019}]{2019MNRAS.487L..47V}
{Vincenzo} F.,  {Spitoni} E.,  {Calura} F.,  {Matteucci} F.,  {Silva Aguirre}
  V.,  {Miglio} A.,   {Cescutti} G.,  2019, \mn@doi [\mnras]
  {10.1093/mnrasl/slz070}, \href
  {https://ui.adsabs.harvard.edu/abs/2019MNRAS.487L..47V} {487, L47}

\bibitem[\protect\citeauthoryear{{Watkins} et~al.,}{{Watkins}
  et~al.}{2009}]{2009MNRAS.398.1757W}
{Watkins} L.~L.,  et~al., 2009, \mn@doi [\mnras]
  {10.1111/j.1365-2966.2009.15242.x}, \href
  {https://ui.adsabs.harvard.edu/abs/2009MNRAS.398.1757W} {398, 1757}

\bibitem[\protect\citeauthoryear{{Wilson} et~al.,}{{Wilson}
  et~al.}{2019}]{2019arXiv190200928W}
{Wilson} J.~C.,  et~al., 2019, arXiv e-prints, \href
  {https://ui.adsabs.harvard.edu/\#abs/2019arXiv190200928W} {p.
  arXiv:1902.00928}

\bibitem[\protect\citeauthoryear{{Xue}, {Rix}, {Ma}, {Morrison}, {Bovy},
  {Sesar}  \& {Janesh}}{{Xue} et~al.}{2015}]{2015ApJ...809..144X}
{Xue} X.-X.,  {Rix} H.-W.,  {Ma} Z.,  {Morrison} H.,  {Bovy} J.,  {Sesar} B.,
  {Janesh} W.,  2015, \mn@doi [\apj] {10.1088/0004-637X/809/2/144}, \href
  {https://ui.adsabs.harvard.edu/abs/2015ApJ...809..144X} {809, 144}

\bibitem[\protect\citeauthoryear{{Yanny} et~al.,}{{Yanny}
  et~al.}{2000}]{2000ApJ...540..825Y}
{Yanny} B.,  et~al., 2000, \mn@doi [\apj] {10.1086/309386}, \href
  {https://ui.adsabs.harvard.edu/abs/2000ApJ...540..825Y} {540, 825}

\bibitem[\protect\citeauthoryear{{Zamora} et~al.,}{{Zamora}
  et~al.}{2015}]{2015AJ....149..181Z}
{Zamora} O.,  et~al., 2015, \mn@doi [\aj] {10.1088/0004-6256/149/6/181}, \href
  {http://adsabs.harvard.edu/abs/2015AJ....149..181Z} {149, 181}

\bibitem[\protect\citeauthoryear{{Zasowski} et~al.,}{{Zasowski}
  et~al.}{2013}]{2013AJ....146...81Z}
{Zasowski} G.,  et~al., 2013, \mn@doi [\aj] {10.1088/0004-6256/146/4/81}, \href
  {http://adsabs.harvard.edu/abs/2013AJ....146...81Z} {146, 81}

\bibitem[\protect\citeauthoryear{{Zasowski} et~al.,}{{Zasowski}
  et~al.}{2017}]{2017AJ....154..198Z}
{Zasowski} G.,  et~al., 2017, \mn@doi [\aj] {10.3847/1538-3881/aa8df9}, \href
  {http://adsabs.harvard.edu/abs/2017AJ....154..198Z} {154, 198}

\bibitem[\protect\citeauthoryear{{de Jong}, {Yanny}, {Rix}, {Dolphin}, {Martin}
   \& {Beers}}{{de Jong} et~al.}{2010}]{2010ApJ...714..663D}
{de Jong} J. T.~A.,  {Yanny} B.,  {Rix} H.-W.,  {Dolphin} A.~E.,  {Martin}
  N.~F.,   {Beers} T.~C.,  2010, \mn@doi [\apj] {10.1088/0004-637X/714/1/663},
  \href {https://ui.adsabs.harvard.edu/abs/2010ApJ...714..663D} {714, 663}

\bibitem[\protect\citeauthoryear{van~der Walt, Colbert  \& Varoquaux}{van~der
  Walt et~al.}{2011}]{5725236}
van~der Walt S.,  Colbert S.~C.,   Varoquaux G.,  2011, \mn@doi [Computing in
  Science Engineering] {10.1109/MCSE.2011.37}, 13, 22

\makeatother
\end{thebibliography}




\appendix

\section{The APOGEE-2 Selection Function}\label{sec:appA}

We build upon the selection function for APOGEE-1 \citep[originally described in][]{2014ApJ...790..127B} to account for the slight adjustments in the targeting scheme which were made between APOGEE-1 and 2. Our description here mainly focuses on the implied selection function due to the finite target selection from the 2MASS catalogue of the APOGEE survey. 

The target selection procedure for APOGEE-2 is discussed in detail by \citet{2017AJ....154..198Z}, and was adjusted slightly from that of APOGEE-1 \citep[][]{2013AJ....146...81Z}. Aside from stars which were targeted specially (e.g. for anciliiary science proposals), APOGEE is generally made from stars which are selected randomly from the 2MASS catalogue within bins in $H$ band apparent magnitude and the dereddened $(J-K_S)_0$ colour. Stars which are selected in this way, and are part of a `completed' observation cohort (discussed below), form part of what is termed the \emph{statistical} sample - on which we perform the density modelling in this paper.

While the exact observing procedure for APOGEE is complex \citep[thorough descriptions can be found in][]{2013AJ....146...81Z,2014ApJ...790..127B,2017AJ....154..198Z}, the procedure for finding the APOGEE selection function boils down simply to assessing for which fields and which colour and magnitude bins observations are completed (at the time the selection function is being assessed), and then simply comparing the number of stars in the spectroscopic, \emph{statistical} sample in that bin to that in the photometric sample. This means that the selection function is a 2D piecewise constant function in colour and magnitude in any given field.

Figure \ref{fig:combsf_cmd} \citep[following Figure 11 of][]{2014ApJ...790..127B} shows the colour-magnitude distribution of the photometric sample combined across all APOGEE-1 and 2 fields, colour and magnitude bins, and demonstrates the quality of our derived selection function. The grayscale histogram shows the density of the photometric sample in APOGEE fields, which is also traced by the black contours. The red contour shows the density of the APOGEE DR14 spectroscopic statistical sample. We generate the blue contour by re-weighting the photometric sample by the selection function at each location, colour and magnitude. The good match between the red and blue contours demonstrates the quality of the correction that can be attained through application of the derived selection function. 

We show the dependence of the combined APOGEE-1 and 2 selection function on sky position in Figure \ref{fig:combsf_lb}. Each row shows a different cohort (corresponding in general to magnitude bins, with long cohorts being those going to the faintest magnitudes), and each column shows a different colour selection, with the blue end of the selection indicated above. The colour limits are defined by only the blue end, as in some cases this is the only limit implied, whereas in others a red limit is also implied (e.g. when a field has two colour bins). Each point in Figure \ref{fig:combsf_lb} represents an APOGEE field, and its colour indicates the derived selection fraction. It is clear that the selection fraction is lower in fields where the stellar density is higher, such as those in the disc plane, and is higher in fields where density and dust extinction are low (i.e. those pointing out of the plane). The selection fraction is high for blue halo fields, and in fields toward the galactic anticenter with a blue colour limit. From this we ascertain that the derived selection function is acting sensibly with sky location.

\begin{figure}
\includegraphics[width=\columnwidth]{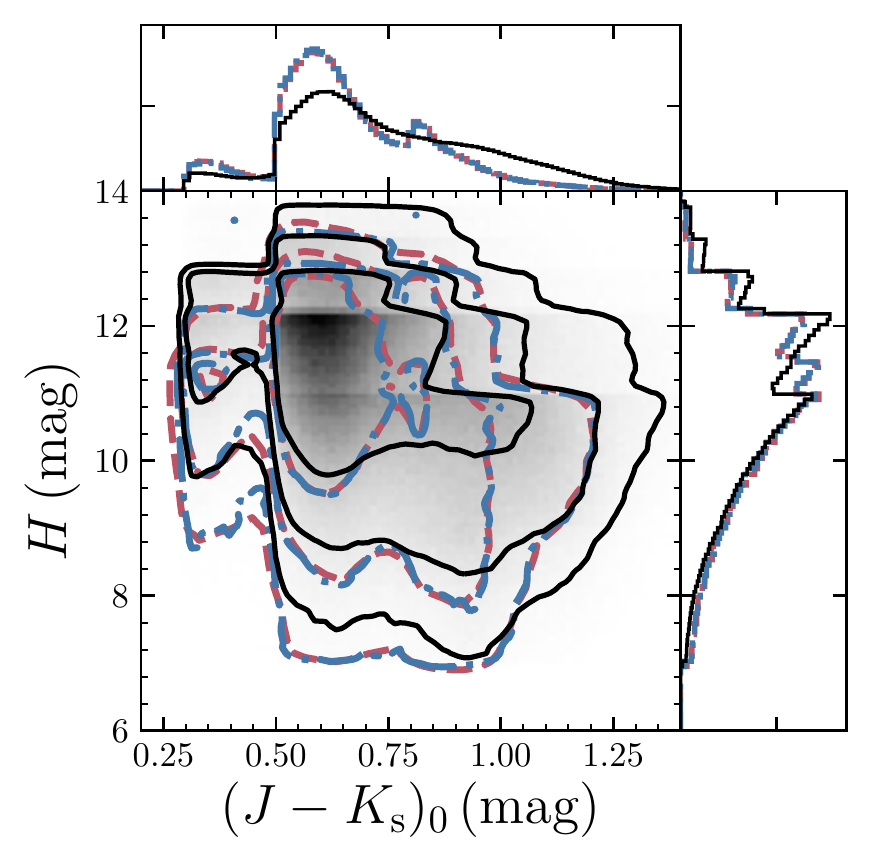}
\caption{\label{fig:combsf_cmd} Comparison of the photometric data in APOGEE fields with the spectroscopic sample drawn from it, in colour-magnitude space. The histogram and black contours show the photometric data taken from 2MASS \citep{2006AJ....131.1163S}. The red contour demonstrates the $(J-K_S)_0$-$H$ distribution of the APOGEE-2 spectroscopic sample, whereas the blue contour shows the same distribution for the photometric data, reweighted by our APOGEE-2 selection function. As the two coloured contours agree well, the selection function is performing correctly in correcting between the spectroscopic number counts and the underlying photometric sample.}
\end{figure}

\begin{figure*}
\includegraphics[width=\textwidth]{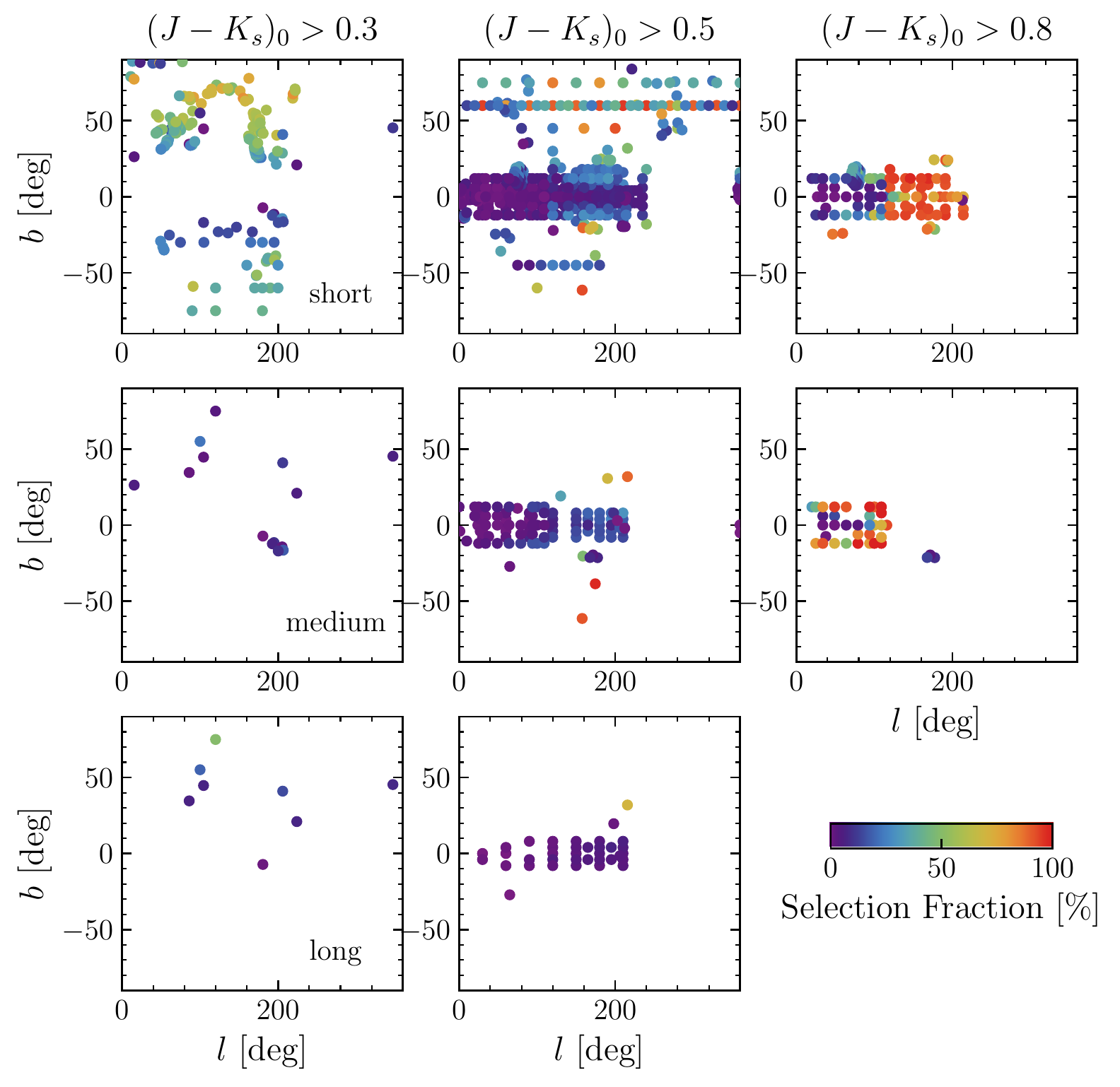}
\caption{\label{fig:combsf_lb} The selection fraction in APOGEE-1 and 2 fields in DR14, as a function of their Galactic latitude and longitude. Each panel shows a different cohort and colour bin, such that the columns show cohort-colour limit combinations with the same colour limit, and rows show combinations in the same cohort. There are no long $(J-K_s)_0 > 0.8$ cohorts. It is clear that in high latitude fields, where a bluer colour limit is adopted, the selection fraction is higher. Similarly, toward the Galactic anticentre, the redder colour limit induces a higher selection fraction. Fields in the disc plane tend to have the lowest selection fraction, owing to the higher stellar density in their pointings.}
\end{figure*}

\section{Testing the application of stellar evolution models with HST photometry}
\label{sec:hst}

In order to verify and test that our usage of stellar evolution models to convert red giant star number counts into the underlying stellar mass of the total population, we use Hubble Space Telescope (HST) UV photometry of the Galactic Globular Cluster (GC) NGC2419. NGC2419 is a good cluster for this purpose, as it is expected to have a near-primordial mass function (i.e. it has undergone little mass segregation), according to the measurement of its mass function slope provided in \citet{2018MNRAS.478.1520B}. Since we na\"ively expect that HST UV photometry should provide complete number counts of stars down to relatively faint magnitudes, we expect that comparing some number of red giant stars (selected in a similar way as APOGEE) to a selection of stars on the main sequence will provide the correct conversion factor between giant star counts to dwarfs, which provide a much more direct tracer of the total mass in a stellar population. By simply checking this number against that we can obtain from the PARSEC isochrones which we adopt, we can ascertain whether the isochrones provide an adequate modelling of the true stellar mass distribution in stellar parameter space, which we rely on for our mass estimates.

\begin{figure}
\includegraphics[width=\columnwidth]{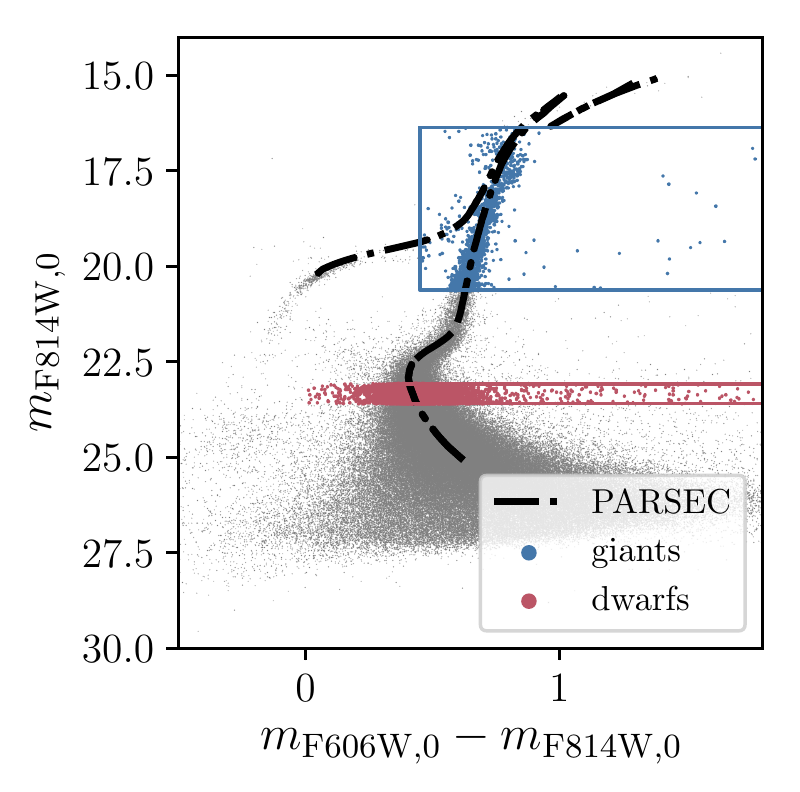}
\caption{\label{fig:ngccmd} The HST WFC3 colour-magnitude diagram of NGC2419, as provided by \citet{2012MNRAS.423..844B}. The PARSEC isochrone corresponding to the age and metallicity of the cluster listed in \citet{2010MNRAS.404.1203F} is overlaid as the dot-dashed line. The red and blue boxes demonstrate the limits we adopt for colour and magnitude selections for dwarf and giant stars to perform our estimate of the mass correction factor.}
\end{figure}

We take HST WFC3 photometry of NGC2419, originally presented in \citet{2012MNRAS.423..844B}, and use this to make our assessment. Figure \ref{fig:ngccmd} shows the WFC3 colour-magnitude diagram for the cluster from the catalogue provided. The corresponding PARSEC isochrone, adopting the age and metallicity for the cluster (age $= 12.3$ Gyr, $\mathrm{[Fe/H]} = -2.2$) listed in \citet{2010MNRAS.404.1203F}, is shown as a dot-dashed line. Each evolutionary stage modelled in PARSEC is represented by a separate line. We adopt the distance to the cluster of $82.6$ kpc and the reddening of $E(B-V) = 0.08$ provided by \citet{1996AJ....112.1487H}, to put the isochrone into the apparent magnitude space ofthe photometry. We select giants as closely as possible to the selection in our APOGEE sample by finding the HST colour which corresponds to a $(J-K_S)_0 > 0.3$ cut, $(m_{\mathrm{F606W}}-m_{\mathrm{F814W}})_0 > 0.451$, then find the magnitude limits in the isochrone that correspond to the limits we impose in $\log(g)$ between 1 and 3, $16.4 < m_{\mathrm{F814W}} < 20.6$. 

We determine a selection of dwarf stars which is minimally affected by the completeness of the photometry by referring to Figure 5 in \citet{2012MNRAS.423..844B}, which provides completeness curves as a function of absolute magnitude and distance from the cluster center.  The photometry is complete down to $M_{\mathrm{F814W}} = 4.$ (corresponding to $m_{F814W} = 23.6$) outside $r = 50"$, which is approximately equal to the half-mass radius $r_h$ of the cluster. We therefore select dwarfs at a range in magnitude $ 3.5 < M_{\mathrm{F814W}} < 4.0$, corresponding to $23.1 < m_{\mathrm{814W}} < 23.6$. We make a cut in colour such that $(m_{\mathrm{F606W}}-m_{\mathrm{F814W}})_0 > 0.3$, to avoid contamination by stars on the horizontal branch. 

We assess the accuracy of the isochrone mass distribution by comparing the ratio of dwarfs to giants in the phometry $\omega_{\mathrm{phot}} = N_{\mathrm{dwarf}}/N_{\mathrm{giant}}$ to the ratio of the integral under the \citet{2001MNRAS.322..231K} IMF at the mass interval along the isochrone in the regions inside our dwarf and giant cuts $\omega_{\mathrm{iso}} = \Delta N_{\mathrm{dwarf}}/\Delta N_{\mathrm{giant}}$. This provides us with an estimate of the correction factor which is required to correct any total stellar mass derived with the isochrones to the actual stellar mass through the ratio $\omega_{\mathrm{phot}}/\omega_{\mathrm{iso}}$. A factor of unity means no correction is required, and the isochrones provide a realistic model of the mass distribution in colour-magnitude and stellar parameters space, in particular along the red-giant branch, that we use in this work.

\begin{figure}
\includegraphics[width=\columnwidth]{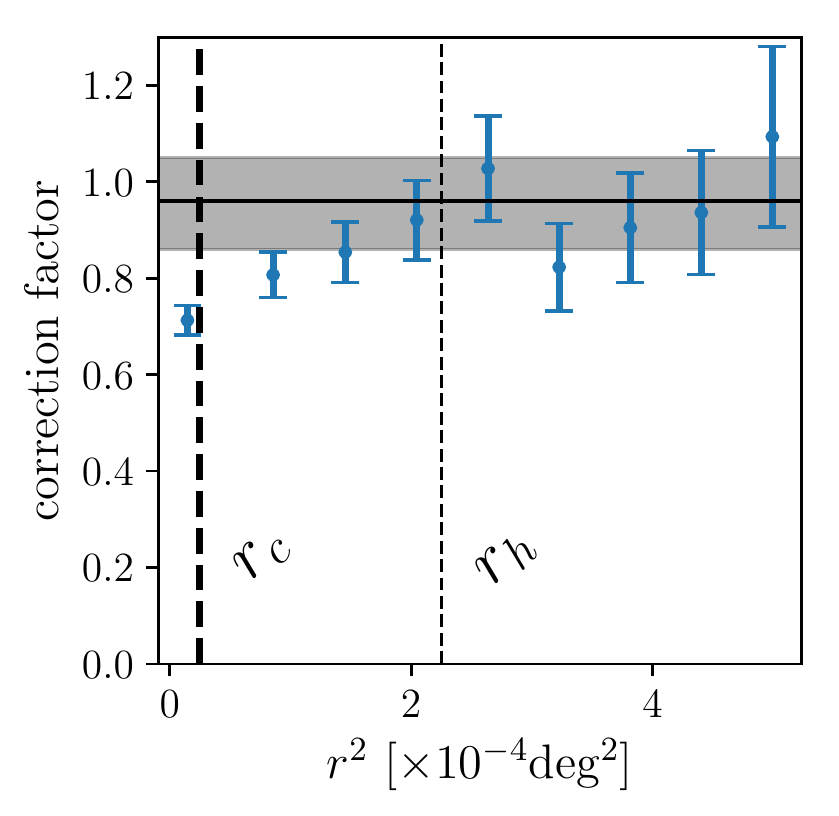}
\caption{\label{fig:ngcrad} The correction factor computed for 9 annuli of equal area extending from the cluster center. We show the factor as a function of the central radius squared for each bin, such that the distance between bins on the x-axis is equal. The core $r_c$ and half-mass $r_h$ radius are shown by annotated dashed lines. We only expect bins outside the $r_h$ of the cluster to provide an evaluation of the correction factor which is not affected by the completeness of the photometry. The mean and standard deviation correction factor outside $r_h$ are demonstrated by the gray band. The mean factor is consistent with being equal to unity.}
\end{figure}

We determine this ratio for a set of annuli of equal area extending from the cluster center, and plot the resulting factors as a function of the radius squared in Figure \ref{fig:ngcrad}. We demonstrate the cluster core $r_c$ and half-mass $r_h$ radius as vertical dashed lines. Within $r_h$, the correction factor is lower than unity, implying that our mass is high by nearly a factor of 2. However, the points outside $r_h$ converge toward unity with some scatter about this value. Since we expect the completeness of the photometry to be significantly reduced inside $r_h$, and likely to be the cause of the slope in the correction factor in this range, we take the mean and standard deviation of the correction factor in the annuli outside $r_h$ and plot this as the gray band. We determine the mean correction factor to be $0.96\pm0.1$, and to therefore be consistent with unity. We therefore can be confident that the mass we derive using the PARSEC isochrones based on red giant star counts is reliable, and does not require correction.

\bsp	
\label{lastpage}
\end{document}